\pdfoutput=1

\documentclass[longauth]{aa}

\usepackage{comment}
\usepackage{graphicx}
\usepackage{multirow}
\usepackage{booktabs}
\usepackage{textgreek}
\usepackage{xargs}
\usepackage[dvipsnames]{xcolor}
\usepackage{xspace}
\usepackage{caption}
\usepackage{subcaption}
\usepackage{float}
\usepackage{enumitem}
\usepackage[rightcaption]{sidecap}
\usepackage{lineno}

\usepackage{txfonts}
\usepackage{hyperref}
\hypersetup{
    colorlinks=true,
    linkcolor=blue,
    citecolor=blue,
    filecolor=magenta,      
    urlcolor=cyan,
    }
\usepackage{subcaption}
\usepackage{adjustbox}

\graphicspath{{./}{figs/}}
\defcitealias{Cataldi2025}{C25}

\newcommandx{\permittedEL}[6][1=O,2=III,3=,4=,5=,6=]{\text{{#1}\,{\sc {#2}}{#3}{#4}{#5}{#6}}\xspace}
\newcommandx{\semiforbiddenEL}[6][1=O,2=III,3=,4=,5=,6=]{\text{{#1}\,{\sc{#2}}]{#3}{#4}{#5}{#6}}\xspace}
\newcommandx{\forbiddenEL}[6][1=O,2=III,3=,4=,5=,6=]{\text{[{#1}\,{\sc{#2}}]{#3}{#4}{#5}{#6}}\xspace}

\newcommand{\Halpha}{\text{H\textalpha}\xspace}
\newcommand{\Hbeta}{\text{H\textbeta}\xspace}
\newcommand{\Hgamma}{\text{H\textgamma}\xspace}
\newcommand{\Hdelta}{\text{H\textdelta}\xspace}

\newcommand{\hii}{\permittedEL[H][ii]}

\newcommandx{\NVL}[1][1=1243]{\permittedEL[N][v][\textlambda][#1]}
\newcommandx{\NVall}{\permittedEL[N][v][\textlambda][\textlambda][1239,][1243]}
\newcommand{\NII}{\forbiddenEL[N][ii]}

\newcommand{\NIV}{\semiforbiddenEL[N][iv]}
\newcommandx{\NIVL}[1][1=1486]{\semiforbiddenEL[N][iv][\textlambda][#1]}

\newcommandx{\CIVL}[1][1=1550]{\permittedEL[C][iv][\textlambda][#1]}

\newcommandx{\HeIIL}[1][1=1640]{\permittedEL[He][ii][\textlambda][#1]}

\newcommand{\OIII}{\semiforbiddenEL[O][iii]}
\newcommandx{\OIIIL}[1][1=1666]{\semiforbiddenEL[O][iii][\textlambda][#1]}

\newcommand{\SiIII}{\semiforbiddenEL[Si][iii]}

\newcommand{\OIIIopt}{\forbiddenEL[O][iii]}
\newcommandx{\OIIIoptL}[1][1=5007]{\forbiddenEL[O][iii][\textlambda][#1]}

\newcommand{\NIII}{\semiforbiddenEL[N][iii]}
\newcommandx{\NIIIL}[1][1=1750]{\semiforbiddenEL[N][iii][\textlambda][#1]}

\newcommandx{\CIII}{\semiforbiddenEL[C][iii]}
\newcommandx{\CIIIL}[1][1=1909]{\semiforbiddenEL[C][iii][\textlambda][#1]}

\newcommandx{\NeIVL}[1][1=2424]{\forbiddenEL[Ne][iv][\textlambda][#1]}

\newcommandx{\MgIIL}[1][1=2803]{\permittedEL[Mg][ii][\textlambda][#1]}

\newcommandx{\NeVL}[1][1=3426]{\forbiddenEL[Ne][v][\textlambda][#1]}

\newcommand{\OII}{\forbiddenEL[O][ii]}
\newcommandx{\OIIL}[1][1=3727]{\forbiddenEL[O][ii][\textlambda][#1]}

\newcommand{\SII}{\forbiddenEL[S][ii]}
\newcommandx{\SIIL}[1][1=6725]{\forbiddenEL[S][ii][\textlambda][#1]}

\newcommandx{\SIIIL}[1][1=9068]{\forbiddenEL[S][iii][\textlambda][#1]}

\newcommand{\NeIII}{\forbiddenEL[Ne][iii]}
\newcommandx{\NeIIIL}[1][1=3869]{\forbiddenEL[Ne][iii][\textlambda][#1]}

\usepackage{etoolbox}
\makeatletter
\newcommand\sendemail[3]{%
\edef\@tempa{mailto:#1?subject=#2 }%
\edef\@tempb{\expandafter\html@spaces\@tempa\@empty}%
\href{\@tempb}{#3}}

\catcode\%=11
\def\html@spaces#1 #2{#1}\catcode\%=14
\makeatother

\let\oldtextsigma\textsigma
\renewcommand{\textsigma}{\oldtextsigma\xspace}
\let\oldtextalpha\textalpha
\renewcommand{\textalpha}{\oldtextalpha\xspace}
\let\oldAA\AA
\renewcommand{\AA}{\text{\oldAA}\xspace}
\let\oldtextdegree\textdegree
\renewcommand{\textdegree}{\oldtextdegree\xspace}

\newcommandx{\Mout}[2][1=,2=]{\ensuremath{M_{\mathrm{out}{#2}}^{#1}}\xspace}
\newcommandx{\Mdotout}[2][1=,2=]{\ensuremath{\dot{M}_{\mathrm{out}{#2}}^{#1}}\xspace}

\newcommandx{\fluxdcgs}[1][1=-20]{\ensuremath{\mathrm{10^{#1}~erg~s^{-1}~cm^{-2}~\AA^{-1}}}\xspace}
\newcommandx{\fluxcgs}[1][1=-20]{\ensuremath{\mathrm{10^{#1}~erg~s^{-1}~cm^{-2}}}\xspace}
\newcommandx{\powercgs}[1][1=44]{$\times 10^{#1}$~erg~s$^{-1}$\xspace}

\begin{document}


\title{Tracing nitrogen enrichment across cosmic time with JWST}

\titlerunning{Nitrogen Enrichment across Cosmic Time}

\author{
E. Cataldi\inst{\ref{UNIFI}, \ref{arcetri}} \and
F. Belfiore\inst{\ref{arcetri}, \ref{eso}} \and
M. Curti\inst{\ref{eso}} \and
B. Moreschini\inst{\ref{UNIFI},\ref{arcetri}} \and
A. Marconi\inst{\ref{UNIFI},\ref{arcetri}} \and
R. Maiolino\inst{\ref{cavensish}, \ref{kicc}} \and
A. Feltre\inst{\ref{arcetri}} \and
M. Ginolfi\inst{\ref{UNIFI},\ref{arcetri}} \and
F. Mannucci\inst{\ref{arcetri}} \and
G. Cresci\inst{\ref{arcetri}} \and
X. Ji\inst{\ref{cavensish}} \and
A. Amiri\inst{\ref{IPM},\ref{arkansas}} \and
M. Arnaboldi\inst{\ref{eso}} \and
E. Bertola\inst{\ref{arcetri}}\and
C. Bracci\inst{\ref{UNIFI},\ref{arcetri}}\and
M. Ceci\inst{\ref{UNIFI},\ref{arcetri}}\and
A. Chakraborty\inst{\ref{arcetri}}\and
F. Cullen\inst{\ref{ROE}} \and
Q. D'Amato\inst{\ref{arcetri}} \and
C. Kobayashi\inst{\ref{herts}}\and
I. Lamperti\inst{\ref{UNIFI},\ref{arcetri}}\and
C. Marconcini\inst{\ref{UNIFI},\ref{arcetri}}\and
M. Scialpi\inst{\ref{trento},\ref{arcetri},\ref{UNIFI}}\and
L. Ulivi\inst{\ref{trento},\ref{arcetri},\ref{UNIFI}} \and
M. V. Zanchettin\inst{\ref{arcetri}}
}

\institute{
\label{UNIFI} Università di Firenze, Dipartimento di Fisica e Astronomia, via G. Sansone 1, 50019 Sesto Fiorentino, Florence, Italy \and
\label{arcetri} INAF -- Arcetri Astrophysical Observatory, Largo E. Fermi 5, I-50125, Florence, Italy \and
\label{eso} European Southern Observatory, Karl-Schwarzschild Straße 2, D-85748 Garching bei München, Germany \and
\label{OAS} INAF -- Bologna Astrophysical Observatory, Via Piero Gobetti, 93/3, 40129, Bologna, Italy \and
\label{IPM}
School of Astronomy, Institute for research in fundamental sciences (IPM), Tehran, P.O. Box 19395-5531, Iran
\and
\label{arkansas} Department of Physics, University of Arkansas, 226 Physics Building, 825 West Dickson Street, Fayetteville, AR 72701, USA \and
\label{SNS} Scuola Normale Superiore, Piazza dei Cavalieri 7, I-56126 Pisa, Italy \and
\label{ROE} Institute for Astronomy, University of Edinburgh, Royal Observatory, Edinburgh, EH9 3HJ, UK \and
\label{herts} Centre for Astrophysics Research, Department of Physics, Astronomy and Mathematics, University of Hertfordshire, Hatfield AL10 9AB, UK \and
\label{aura} AURA for European Space Agency, Space Telescope Science Institute, 3700 San Martin Drive. Baltimore, MD, 21210 \and
\label{cavensish} Cavendish Laboratory, University of Cambridge, 19 JJ Thomson Avenue, Cambridge, CB3 0HE, UK \and
\label{kicc} Kavli Institute for Cosmology, University of Cambridge, Madingley Road, Cambridge, CB3 0HA, UK \and 
\label{trento} University of Trento, Via Sommarive 14, I-38123 Trento, Italy
}

\authorrunning{E. Cataldi et al.}
\date{}

\abstract{
We present a comprehensive analysis of the nitrogen-to-oxygen (N/O) abundance ratio in a sample of $\sim 660$ star-forming galaxies at redshift $z\sim 1-6$, with a median redshift of $\langle z \rangle$ = 3.0, using deep JWST/NIRSpec spectroscopy. 
Leveraging detections of faint auroral emission lines in 92 galaxies at $z > 1$ from both the MARTA survey and a large compilation of high-redshift literature objects, we derive direct electron temperature-based abundances for nitrogen and oxygen using rest-frame optical lines. We establish the first high-redshift calibrations of strong-line N/O diagnostics based on `direct' abundance measurements, finding a mild evolution for the N2O2 = $\log$(\NII $\lambda$6585/\OII $\lambda\lambda$3727,3729) diagnostic and no clear evolution for the N2S2 = log(\NII$\lambda$6585/\SII$\lambda\lambda$6717,6731) diagnostic compared to local realisations. We then investigate the N/O–O/H relation across cosmic time using both `direct' abundances and strong-lines based measurements (additional 535 galaxies). We find evidence for mild but systematic nitrogen enhancement at high redshift: galaxies at $z > 1$ exhibit N/O ratios elevated by $\sim$0.18 dex (median offset) at fixed O/H compared to local trends, with a more pronounced enhancement at low metallicity (i.e. 12 + log(O/H) $\lesssim$ 8.1), where the offset reaches up to $\sim$0.4-0.5~dex. We consider several scenarios to explain the observed trends, including bursty star formation, differential metal loading, and inflows of pristine gas. Our results provide the most extensive confirmation of elevated N/O ratios at high-redshift to date based on rest-optical diagnostics and within a self-consistent frame. 
}


   \maketitle
   
\nolinenumbers 

\section{Introduction}

Metals are essential tracers of galaxy assembly, encoding the imprint of star formation, gas accretion, and feedback processes across cosmic time \citep[e.g.,][]{maiolino_re_2019}. Their presence regulates key processes in the interstellar medium (ISM), as they set the efficiency of radiative cooling \citep[e.g.,][]{sutherland_cooling_1993}, modulate the conversion of gas into stars \citep[e.g.,][]{krumholtz_2012, bolatto_co--h2_2013}, and determine the opacity and thermodynamic state of the ISM \citep[e.g.,][]{Draine_2011}. Metals are therefore central actors in the baryon cycle: constraining the timing and efficiency of metal production and tracing the evolution of their abundances with redshift is therefore essential for reconstructing the pathways of galaxy formation and evolution.

Beyond absolute abundances, the relative abundance patterns of individual elements encode information on the timescales and sources of chemical enrichment, and are therefore powerful diagnostics of the physical processes driving galaxy evolution. Among these, the nitrogen-to-oxygen ratio (N/O) plays a particularly important role. In fact, oxygen is an $\alpha$-element, produced predominantly through helium fusion in short-lived massive stars ($M \gtrsim 8,M_\odot$), which end their lives as core-collapse supernovae, enriching the ISM with oxygen.
Nitrogen, on the other hand, can be produced through two main channels, commonly referred to as primary and secondary production. In the primary channel, nitrogen originates from carbon and oxygen freshly synthesized within the star itself: these newly formed nuclei are mixed into the hydrogen-burning regions and converted into nitrogen through the CNO cycle. As a result, the nitrogen yield in this case is largely independent of the star’s initial metallicity. Primary nitrogen can be produced both by massive stars on short timescales--tens of Myr--where fast rotational and convective mixing processes are highly efficient \citep[e.g.,][]{2000ApJ...541..660H,  2002A&A...390..561M,chiappini_stellar_rotation_2006}, and by low- and intermediate-mass stars (LIMS hereafter; $M\lesssim8 M_\odot$), during their asymptotic giant branch (AGB) phase \citep[e.g.,][]{vincenzo_nitrogen_2016}. Such production channel happens on much longer enrichment timescales--hundreds of Myr to several Gyr. The relative importance of each channel depends on the star formation history, IMF, and metallicity evolution of the galaxy \citep[e.g.,][]{Romano_2019}. In contrast, the secondary channel relies on carbon and oxygen already present in the progenitor gas cloud. In this case, the CNO cycle transforms pre-existing heavy elements into nitrogen, leading to a metallicity-dependent yield \citep[e.g.,][]{2000ApJ...541..660H, vincenzo_nitrogen_2016, maiolino_re_2019}. This contribution arises mainly from LIMS in their AGB phase \citep{kobayashi_isotopes_MW_2011, nomoto_chemical_evolution_2013, ventura_agb_yields_2013}.

As a consequence, the N/O abundance ratio provides crucial insights into the timing and efficiency of chemical enrichment processes driving galaxy evolution \citep[e.g.,][]{2018MNRAS.477...56V, vincenzo_NO_simulations_2018}. In fact, the detailed shape and normalization of the N/O-O/H relation, as well as its scatter, are influenced by several factors, including the star-formation history, the initial mass function, metallicity-dependent stellar yields, gas inflows and outflows, in addition to the time delay between oxygen enrichment from massive stars and nitrogen release from LIMS \citep[e.g.,][]{1980FCPh....5..287T, 1990MNRAS.246..678E, 1992A&A...264..105M, 2002A&A...390..561M,  ventura_agb_yields_2013, vincenzo_nitrogen_2016, 2000ApJ...541..660H, Arellano-Cordova25b}. 

In the local Universe, observations based on the direct method (the so-called $T_e$ method) reveal a well-defined N/O–O/H relation. N/O remains approximately constant at low oxygen abundance, consistent with primary nitrogen production being the dominant process, and then rises steeply above $12 + \log(\mathrm{O/H}) \gtrsim 8.3$, where secondary nitrogen from LIMS becomes significant \citep[e.g.,][]{garnett_nitrogen_1990,vila_costas_nitrogen-ratio_1993, 2006MNRAS.372.1069M, berg_direct_2012, pilyugin_metallicity_2014, belfiore_p-manga_2015, vincenzo_nitrogen_2016, vincenzo_extragalactic_2018}. This trend is observed both in individual \hii regions and in integrated galaxy spectra, and it broadly agrees with predictions of chemical evolution models \citep[e.g.,][]{matteucci_considerations_1986, molla_grid_2005, vincenzo_extragalactic_2018}.

Direct determination of nitrogen and oxygen abundances at high redshifts have long been limited by observational challenges: the rest-frame optical lines that enable such measurements are shifted into the near-infrared at $z \gtrsim 2$, and the required auroral lines are intrinsically faint.

Analyses based on strong-line ratios that correlate with N/O, such as N2O2 - defined as $\log(\NII~\lambda6584/ \OII~\lambda\lambda3727,3729 )$ - and N2S2 - defined as $\log(\NII~\lambda6584/ \SII~\lambda\lambda6717,6731 )$ - generally find elevated N/O at fixed metallicity relative to $z\sim0$ samples \citep[e.g.,][]{masters_physical_2014, strom_nebular_2017, hayden-pawson_NO_2022}
The fundamental limitation of these efforts is that strong-line diagnostics are inherently sensitive to additional parameters, especially the ionisation parameter and abundance ratios like S/O \citep[e.g.,][]{dopita_chemical_2016, 10.1093/mnras/stab862, 10.1093/mnras/staf834}. Without detections of faint auroral lines and corresponding electron temperature estimates, it has remained unclear whether observed trends reflect genuine chemical differences or evolving ISM conditions. 

The sensitivity and wavelength coverage of JWST/NIRSpec \citep{jakobsen_nirspec_2022} now allow robust abundance determinations based on auroral-line detections in galaxies out to $z \gtrsim 10$ \citep[e.g.,][]{bunker_gnz11_2023, cameron_gnz11_2023, Castellano2024, marques-chaves_Nitrogen_2024,  2024A&A...687L..11S, Topping2024, 2025A&A...697A..89C, Napolitano_2025, Navarro_Carrera25}.

At intermediate redshift ($z \sim 2$), these new observational capabilities open the possibility of reassessing the N/O–O/H relation with unprecedented precision \citep{Sanders_2023,Strom2023CECILIA:, 2024ApJ...964L..12R, 2024ApJ...975..196W, 2025MNRAS.540.2991A, 2025arXiv250906622C, Sanders_2025, 10.1093/mnras/staf834, 2025MNRAS.541.1707T, 2025ApJ...980...33W, Rogers2026}.

At even earlier epochs, JWST spectroscopy has recently uncovered surprising evidence for extreme nitrogen enrichment, indicative of super-solar N/O ratios at sub-solar O/H, well above the expectations from standard chemical evolution and nucleosynthetic timescales 
\citep{bunker_gnz11_2023, cameron_gnz11_2023, isobe_CNO_2023, jones_CO_z6_2023, Castellano2024, 2024MNRAS.535..881J, 2024A&A...687L..11S,  Topping2024, 2025A&A...697A..89C, Napolitano_2025,   Navarro_Carrera25,  2025ApJ...980..225T,    2025ApJ...981..136S, Ji2026, Zhang2026}. Standard chemical evolution models cannot fully account for this nitrogen enhancement. Proposed alternative channels include enrichment by fast-rotating low-metallicity stars sampled from a top-heavy IMF, very massive stars (VMS; $M_*>100M_{\odot}$), Wolf-Rayet stars, early AGB enrichment combined with gas flows, bursty star formation history, pristine gas inflows, differential outflows, and rare events such as tidal disruptions 
\citep{2023A&A...679L...9V, charbonnel_gnz11_2023, 2023A&A...680L..19D, Nagele_Umeda_N_gnz11_2023, watanabe_2024, 2024MNRAS.533..687R, Arellano-Cordova25b, 2025A&A...697A..96R, Bhattacharya26, McClymont2026}.

In addition to individual detections, stacking analyses also suggest that elevated N/O may be relatively common at $z>6$: for example, \citet{2025ApJ...982...14H} report enhanced N/O in composite spectra spanning $4<z<10$. The agreement between N/O values inferred from stacked spectra and from single galaxies indicates that nitrogen enrichment may not be restricted to rare objects, but could instead reflect conditions that were widespread in the early Universe. 

However, it is important to note that both these individual and stacked measurements of nitrogen enhancements obtained so far with JWST have been derived from rest-frame UV line ratios, such as \NIII$\lambda\lambda$1747,1749, \NIV$\lambda$1486, and \OIII$\lambda$1666. These lines generally probe different, denser, and higher ionisation regions from those probed by rest-frame optical ratios. This effect may introduce systematic differences when compared to optical abundance diagnostics \citep[e.g.][]{Acharyya2019, Byler2020, Mingozzi2022CLASSY, Martinez2025}.
Moreover, pre-JWST studies could be subject to strong selection biases, as ground-based detections of nitrogen lines at high $z$ require extremely bright, massive, and luminous galaxies. This observational bias favours systems with intrinsically bright nitrogen line emission, making it difficult to assess whether such N/O enhancement is representative of the general galaxy population.
By contrast, recent JWST measurements using rest-frame optical nitrogen lines at $z \lesssim 4.5$ find either no significant deviation from the local and lower-redshift N/O–O/H relations  \citep{Arellano-Cordova25b, 2025ApJ...981..136S, Rogers2026} or mild enhancement \citep{2025MNRAS.540.2991A}, except in the case of clearly exceptional systems such as the Sunburst Arc \citep{2025ApJ...980...33W}. 

This picture motivates a renewed focus on N/O as a tracer of chemical enrichment and baryon cycling in galaxies. By leveraging auroral-line measurements with JWST at Cosmic Noon ($z\sim2-3$, the peak epoch of star formation) this work presents the first calibration of N/O–strong line relations fully anchored at high redshift, and the first comprehensive characterization of the N/O–O/H relation based entirely on high-$z$ galaxies. In particular, we combine direct auroral-line abundance determinations and strong-line estimates derived from high-redshift calibrations, thus ensuring a consistent framework across a sample of 373 galaxies at $z>1$. This approach extends beyond previous studies, which relied exclusively on local calibrations, providing unprecedented insights into the mechanisms of nitrogen enrichment during the peak epoch of galaxy formation.

The paper is organized as follows. In Section~\ref{sec:data} we present the extensive high-redshift and local comparison samples used for our abundance analysis. Section~\ref{sec:ISM_prop_sec_3} details the methodology adopted to derive direct, $T_{\rm e}$-based abundances and to estimate them from strong-line diagnostics when auroral lines are unavailable. 
In Section~\ref{sec:results} we recalibrate the N2O2–N/O and the N2S2-N/O relations for high-$z$ galaxy samples, using galaxies with direct measurements. We then analyse the intrinsic scatter and physical drivers of these relations, and finally investigate the N/O–O/H relation across the combined sample. Sec. \ref{sec:discussion} discusses of how our results fit into the recent literature on nitrogen enrichment at high redshift. Our main conclusions are summarized in Section~\ref{sec:conclusions}.

\section{Data}
\label{sec:data}
We aim to select the largest possible sample of high-redshift galaxies for which reliable abundance can be measured, either with the direct method or using strong line calibrations tailored to the high-redshift ISM conditions.
For comparison, we also assembled an extensive compilation of local galaxies and \hii\ region spectra, selected according to the same criteria. 
In particular, in Section~\ref{sec:auroral_data} we describe the selection of galaxies with auroral-line detections, including both high-redshift objects observed with JWST, as well as local galaxies with auroral-line detections enabling direct metallicity measurements. In Section~\ref{sec:strong_line_data}, we discuss the selection of galaxies where we use strong-line abundance diagnostics, comprising several ground-based spectroscopic surveys targeting Cosmic Noon, a large compilation of archival data observed with JWST, and local galaxies from large spectroscopic surveys. 

Across all datasets presented in this work, we applied consistent selection criteria based on emission line signal-to-noise ratios. Specifically, 
we required S/N $\ge$ 5 for the Balmer lines used in the dust correction (primarily \Halpha and \Hbeta, but also \Hgamma and \Hdelta when available), and S/N $\ge$ 3 for the lines used in the abundance diagnostics, namely \OII and \NII. For datasets including auroral line detections, the auroral lines were also required to have S/N $\ge$ 3. When the \SII doublet was used in the analysis, it was similarly selected with S/N $\ge$ 3. Finally, we adopted standard diagnostics from \citet{kewley_cosmic_2013} to exclude potential active galactic nuclei (AGN) contamination.
The redshift distribution of the high-redshift sample is shown in Figure \ref{fig:z_hist}, demonstrating that the bulk of our objects lie between $z \sim 2$ and $z \sim 4$ (median redshift $z=3.0$).

\subsection{Auroral-line sample}
\label{sec:auroral_data}
\subsubsection{The MARTA survey}

The MARTA (Measuring Abundances at high Redshift with the T$_e$ Approach; PID 1879, PI Curti) programme targets 126 star-forming galaxies at Cosmic Noon ($z \sim 2 - 3$), leveraging ultra-deep, medium-resolution JWST/NIRSpec spectroscopy to probe the physical conditions of the ISM. The primary goal is the detection of several rest-frame optical auroral lines -- most notably \OIIIopt$\lambda4363$ using the G140M grating (R$\sim$1000) and \OII$\lambda\lambda$7320,7330 with G235M grating -- to enable direct T$_e$ measurements and reliable gas-phase metallicity determinations. The construction of the parent catalogue, target selection, and prioritization strategy are described in detail in 
\citealt[][hereafter \citetalias{Cataldi2025}]{Cataldi2025}. Observations comprise a total integration time of 31.9 hours with the G140M grating and 7.4 hours with G235M. 
Details on the spectral properties, fitting strategy, and validation are provided in \citetalias{Cataldi2025}. 
The S/N selection criteria described above yield a total of 27 galaxies meeting the requirements. Among these, ten MARTA galaxies also meet the requirements of the auroral-lines sample, allowing for direct determinations of the oxygen and nitrogen abundances.

\begin{figure}
    \centering
    \includegraphics[width=\linewidth]{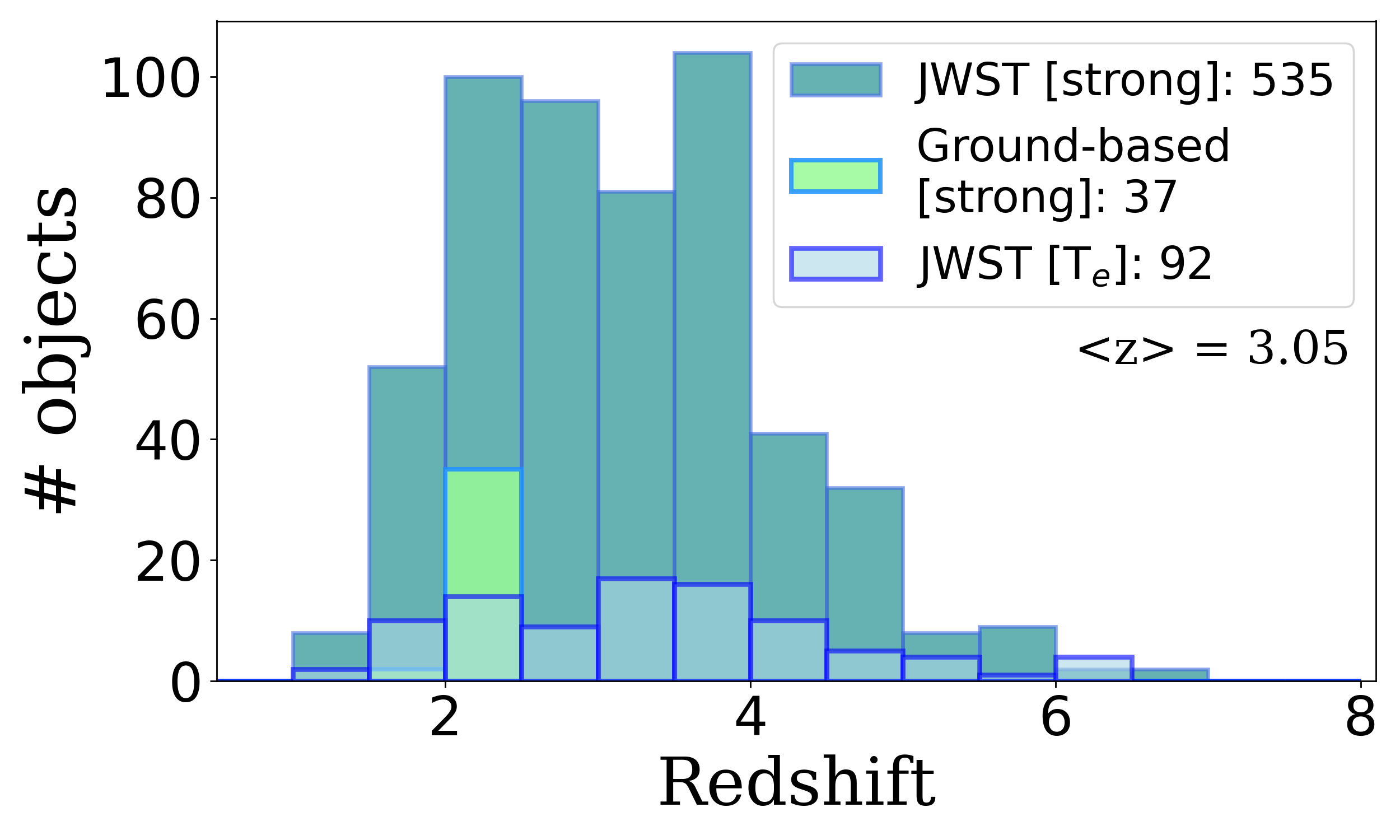}
    
    \caption{Redshift distribution of the 664 galaxies for which we measure O/H and N/O in the high-$z$ sample. JWST galaxies with auroral-line detections are shown in light-blue; JWST objects without auroral-line detections (strong-line measurements only) are displayed in teal; objects from ground-based surveys (strong-line measurements from MOSDEF and KLEVER) are plotted in green.
    }
    \label{fig:z_hist}
\end{figure}

\subsubsection{JWST archival sample}
\label{sec:high_z}

We compiled a sample of 11 galaxies at $z>$1 observed with JWST from the literature meeting all the selection criteria required for auroral-lines sample.
Two of these objects also exhibits detections of the \OII$\lambda\lambda$7320,7330 auroral doublet. An additional 66 galaxies meeting the same selection criteria were drawn from version 4 of the DAWN JWST Archive \citep[DJA; \doi{10.5281/zenodo.1547235}][]{2025A&A...699A.358V}, a public repository containing reduced spectroscopic data from archival JWST observations processed with MSAExp \citep{2022zndo...7299500B}. Emission line fluxes for these galaxies were adopted directly from the standard DJA catalogue. 
The sample selection was based on the detection of optical \NII lines, while nitrogen UV lines were not considered, to avoid potential systematic effects due to the different tracers.

\subsubsection{Local universe sample}  
\label{sec:local_objects}

For comparison with the sample presented in the previous section, we assembled a large sample of low-redshift galaxies and local \hii regions, meeting all the S/N requirements of the auroral-line sample, enabling direct metallicity determinations, for a total of 904 local objects. All abundances were re-derived in a self-consistent manner, by recovering the published line fluxes and re-computing electron temperatures following a uniform methodology. 
The local compilation includes:  

\begin{itemize}
    \item low-metallicity emission-line galaxies and \hii\ regions from \citet{2009A&A...503...61I, 2009A&A...505...63G, guseva_vlt_2011};  
    \item the \citet{curti_new_2017} sample of Sloan Digital Sky Survey (SDSS) Data release 7 spectra,  stacked in bins of \OII/H$\beta$ and \OIIIopt/H$\beta$ line ratios;  
    \item individual SDSS galaxies with auroral-line detections (priv. comm. Amiri);  
    \item \hii regions from the CHAOS (CHemical Abundances Of Spirals) project, targeting nearby spirals with LBT/MODS \citep{berg_chaos_2015, berg_chaos_2020, rogers_chaos_2021, rogers2022chaos};  
    \item the SDSS-IV MaNGA stacks of star-forming galaxies as a function of radius from \citet{Khoram_2025}, which combine spectra across the $M_\star$--SFR plane;  
    \item the compilation by \citet{2021MNRAS.500.2359Z}, consisting of \hii\ region measurements in 51 local spirals from various literature sources;
    \item the BOND (Bayesian Oxygen and Nitrogen abundance Determinations) sample from \citet{vale_asari_bond_2016}, composed of local blue compact dwarfs and giant \hii regions.
    
\end{itemize}

\subsection{Strong-line sample}

\label{sec:strong_line_data}

\subsubsection{$z>1$ sample}
\label{sec:high_z_strong_lines}

We included galaxies from the ground-based MOSDEF \citep[MOSFIRE Deep Evolution Field survey,][]{kriek_mosfire_2015} and KLEVER \citep[KMOS Lensed Emission-Line Velocity and Metallicity Survey,][]{curti_klever_2020} surveys. These surveys do not detect auroral lines but provide all the strong emission line fluxes required to derive metallicities and nitrogen abundances through indirect methods, in particular \OII$\lambda\lambda$3727,3729 and \NII$\lambda$6585. MOSDEF (PI: M. Kiek) is a spectroscopic survey carried out with MOSFIRE \citep{mclean_mosfire_2012} on the 10m Keck I telescope, targeting galaxies at $1.4 \leq z \leq 3.8$. KLEVER, on the other hand, is an ESO Large Programme (PID: 197.A-0717, PI: M. Cirasuolo) that observed 192 galaxies in the redshift range $1.2 < z < 2.5$ with KMOS. We applied the S/N selection criteria described at the beginning of this section.

We also considered galaxies from the public release of the JWST Advanced Deep Extragalactic Survey (JADES) Data Release 4 \citep{robertson_jades_2022, Eisenstein_JADES_2023, 2025arXiv251001033C, 2025arXiv251001034S}. From this catalogue, we selected sources at $z>1$ without auroral-line detections (as those with detected auroral lines are already included in the high-redshift sample described above), but with robust measurements of the strong nebular lines relevant to our analysis.  
This selection yielded a final sample of 188 galaxies from JADES DR4 suitable for inclusion in our analysis.
Finally, we complemented this dataset with additional galaxies drawn from version 4 of the DAWN JWST Archive selected following the same criteria, as described in Section~\ref{sec:high_z}, but without the auroral-line requirement. This brings the total number of high-redshift, strong-line–based comparison galaxies to 535.
The distribution of our high-redshift galaxies is consistent with the star-forming locus, with no clear evidence for contamination by AGN or LINERs (low-ionisation nuclear emission-Line region).

\subsubsection{Local universe sample}
\label{ref:local_obj_strong_lines}

To provide a comparison with a sample representative of the local galaxy population and extend the local comparison sample to high metallicity we leverage the SDSS Data Release 8 MPA-JHU catalogue \citep{brinchmann_physical_2004}. 
In addition to the S/N selection criteria applied uniformly across all datasets, we restricted this sample to local redshifts ($z<0.27$). The final SDSS sample includes approximately 87,000 galaxies.

\begin{figure}
    \centering
    \includegraphics[width=1\linewidth]{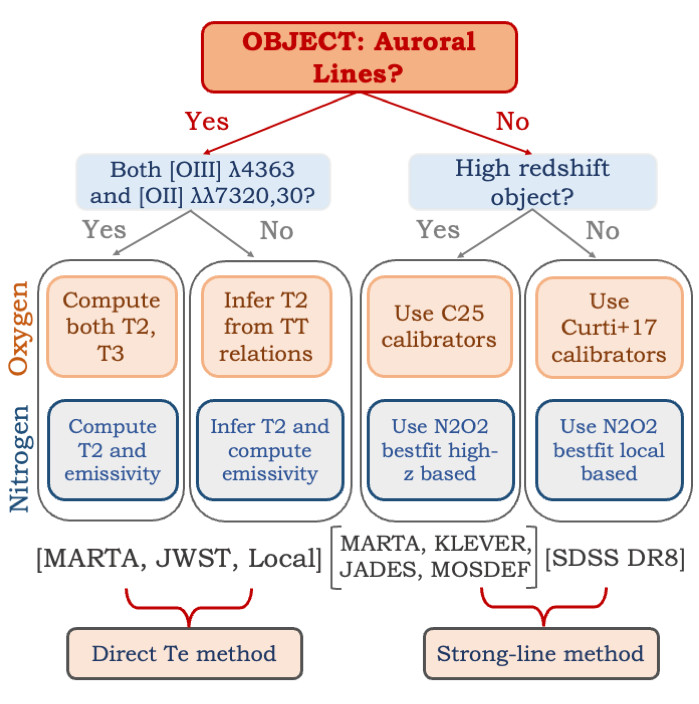}
    \caption{Flowchart summarizing the decision process adopted to compute nitrogen and oxygen abundances depending on the available spectral features. The first step is to assess whether the dataset includes detections of auroral lines. If two oxygen auroral lines are detected, the direct method is applied: both $T_{2}$ and $T_{3}$ temperatures, as well as the electron density, are derived, providing all the quantities needed to compute nitrogen abundances and metallicities. If only one auroral line is available, the corresponding temperature is measured directly, while the second is inferred from a $T_{2}$--$T_{3}$ relation appropriately calibrated --- using a local-based relation for nearby galaxies, or the C25 calibration for high-redshift objects. If no auroral detections are present, the procedure diverges between local and high-redshift samples. For high-$z$ galaxies, metallicities are obtained from strong-line calibrations of C25, and nitrogen abundances from the corresponding high-$z$ N/O--strong-line relations. For local samples, metallicities are derived from the \citet{curti_new_2017} calibrations, while nitrogen abundances follow N/O--strong-line relations calibrated on SDSS local stacks. The Local sample in this figure refers to the dataset described in Section \ref{sec:local_objects} and JWST refers to the Sample depicted in Section \ref{sec:high_z}.}
    
    \label{fig:scheme}
\end{figure}

\section{Determination of ISM properties}
\label{sec:ISM_prop_sec_3}

\subsection{Dust reddening correction}
\label{sec:dust_correction}

For all datasets, emission-line fluxes were corrected for dust attenuation following the same, consistent approach. We first selected the Balmer lines H$\alpha$, H$\beta$, H$\gamma$, and H$\delta$ when available, requiring a S/N > 5. From these, we computed up to three independent Balmer line ratios, assuming the theoretical values for case B recombination at an electron temperature $T_e = 10^4$ K and density $n_e = 10^2$ cm$^{-3}$ \citep{osterbrock_astrophysics_2006}. To assess the impact of our assumptions, we implemented an iterative procedure in which the attenuation correction was recalculated at each step using updated $T_{\rm e}$ and $n_{\rm e}$ values from the previous iteration derived as described in Sec. \ref{sec:met_determination_direct}, until convergence. The resulting variations in $E(B-V)$, $T_{\rm e}$, and $n_{\rm e}$ are minimal -- typically below 2\% across the sample -- indicating that adopting fixed temperatures and densities does not introduce any significant bias. We note that this approach can only be applied to a small subsample of objects ($\sim$60 out of 660) for which both auroral and sulphur lines are detected, as these are required to directly constrain $T_{\rm e}$ and $n_{\rm e}$. Applying the MCMC method only to this limited subset while adopting the standard approach for the rest of the sample would introduce heterogeneity in the analysis. Having verified that the two approaches yield very similar results, we therefore adopt the standard methodology consistently across the full dataset.

To compute $E(B-V)$ we applied the same procedure as in \citetalias{Cataldi2025}, performing a global $\chi^2$ minimization to simultaneously fit the available observed ratios with the adopted extinction curve and derive a single best fit. In all cases, the extinction correction was computed adopting the \citet{cardelli_relationship_1989} extinction law. Employing alternative attenuation laws, such as the \cite{calzetti_dust_1994} or the Small Magellanic Cloud law \citep{gordon_LMC_attenuation_2003}, produces negligible differences.

We also tested the impact of adopting a high-redshift nebular attenuation curve as recently proposed by \citet{Reddy2026}, using their `median' attenuation law (Eq.~22). The two laws yield systematically different $E(B-V)$ values, with the \citet{Reddy2026} curve returning generally lower estimates compared to \citet{cardelli_relationship_1989}, with a median offset of $-0.06$~dex. This translates into typical variations of $\sim +0.03$~dex in N2O2, a key quantity relevant for our conclusions, and of similar order in other properties (e.g., $\sim 50$~K in electron temperatures). Importantly, these differences do not affect our overall conclusions.

\subsection{Oxygen abundances from direct measurements}
\label{sec:met_determination_direct}

Determining the chemical composition of a photoionised nebula requires knowledge of its internal temperature and density structure \citep{stasinska_abundance_2002}. A standard framework assumes that the nebula consists of two principal ionisation regions: a low-ionisation zone and a high-ionisation zone. Within this scheme, the electron temperature associated with O$^{2+}$, denoted $T_3$, traces the conditions in the high-ionisation region, while the temperature linked to O$^{+}$, $T_2$, characterises the low-ionisation zone dominated by species such as S$^{+}$ and N$^{+}$. Intermediate ions, like S$^{2+}$, can straddle both regimes \citep[e.g.][]{berg_chaos_2020}.

We derived gas-phase oxygen abundances using a consistent, multi-tiered approach depending on the availability of auroral emission lines. Figure~\ref{fig:scheme} illustrates how the adopted procedure varies depending on the set of emission lines available - and the redshift regime - across our different datasets. 

In particular, our primary strategy relied on the $T_{\rm e}$ method, whereby T$_e$ and n$_e$ are used to compute ionic abundances. When both the \OIIIopt~$\lambda4363$ and \OII~$\lambda\lambda7320,7330$ auroral lines were detected, we directly measured the electron temperature in both ionisation zones: the low-ionisation temperature ($T_2$) from the \OII~$\lambda\lambda7320,7330$ / \OII~$\lambda\lambda3727,3729$ ratio, and the high-ionisation temperature ($T_3$) from the \OIIIopt~$\lambda4363$ / \OIIIopt~$\lambda5007$ ratio.
Accurate determination of the ionic abundances also requires a diagnostic of the electron density $n_{\rm e}$, based on a density-sensitive doublet. In the majority of our cases, this is provided by the \SII~$\lambda\lambda6716,6731$ doublet, which we use together with the temperature diagnostics.

Specifically, $T_2$ and the electron density $n_{\rm e}$ were derived simultaneously using the \texttt{getCrossTemDen} routine in \texttt{PyNeb} \citep{luridiana_pyneb_2015}, thus accounting for the density dependence of the \OII temperature diagnostic. The density value obtained from this step was then adopted to compute $T_3$ with the \texttt{getTemDen} function, based on the measured \OIIIopt~$\lambda4363$ / \OIIIopt~$\lambda5007$ ratio. We verified that the inferred $T_3$ values are insensitive to the assumed $n_{\rm e}$ within the range 10–1000 cm$^{-3}$, varying by less than 5\%.
The ionic abundances O$^+$ and O$^{++}$ were then derived through the \texttt{getIonAbundance} routine, adopting the corresponding $T_2$, $T_3$, and $n_{\rm e}$ values.
We adopted the same set of atomic data and collisional strengths as described in \citetalias{Cataldi2025}. 

For galaxies in which only one auroral line was detected -- typically \OIIIopt~$\lambda4363$ -- we estimated $T_2$ using the empirical $T_2$–$T_3$ relation derived in \citetalias{Cataldi2025}. This relation, calibrated on galaxies with both auroral lines at Cosmic Noon, allowed us to extend the direct-method metallicity determination to objects with partial auroral coverage.
For a few MARTA and JWST sources with only one auroral line, the \SII doublet was unavailable, either due to limited spectral coverage or insufficient S/N. In those cases we adopted a fiducial value of $n_{\rm e} = 300$\,cm$^{-3}$, consistent with the typical densities measured in high-$z$ galaxies \citep[e.g.][]{sanders_mosdef_2016, Stanton_excels_2025, 2025MNRAS.541.1707T}. We verified that varying $n_{\rm e}$ within a plausible range (10–1000 cm$^{-3}$) has only a minor impact (less than 1\%) on the resulting $T_3$ values and does not affect the derived metallicities.
In those cases in which, instead, the ratio between the two sulphur lines placed the galaxy in the low-density limit, we assumed a fiducial lower limit of $n_e=100$~cm$^{-3}$.

Uncertainties on the derived quantities were assessed through Monte Carlo simulations; the same approach was systematically applied throughout this work to evaluate the uncertainties on all derived parameters. Each observed emission-line flux was perturbed by adding random Gaussian noise based on its measurement error, and the electron temperature and density were recomputed for 300 realisations\footnote{This number ensures convergence, in the sense that the results do not change if using a larger value.}. The adopted values and 1$\sigma$ uncertainties correspond to the mean and standard deviation of these distributions. Owing to the very high S/N of the Balmer lines (typically exceeding 30–300 for \Halpha\ and \Hbeta, and 10–100 for \Hgamma\ and \Hdelta), the resulting variations in $E(B-V)$ are negligible. For this reason, the extinction correction was kept fixed during the Monte Carlo runs, while perturbations were applied only to the oxygen and sulphur lines that directly influence the determination of $T_{\rm e}$ and $n_{\rm e}$.

\subsection{Oxygen abundances from strong lines}
\label{sec:oxygen_from_strong_lines}
In cases where no auroral lines were available we used the set of strong-line metallicity calibrations presented in \citetalias{Cataldi2025} for the high-redshift sample and the calibrations of \cite{curti_new_2017} for the low-redshift one. Specifically, for each galaxy with multiple strong-line detections, we considered a pool of diagnostic ratios and inferred the metallicity by minimizing the $\chi^2$.
We tested various combinations of diagnostics and converged on [R2, R3], where $R2=\log(\OII\lambda\lambda3727,3729/\Hbeta)$ and $R3=\log(\OIIIopt\lambda5007/\Hbeta)$. This pair of diagnostics leads to the lowest $\chi^2$ with respect to the calibrations presented in \citetalias{Cataldi2025}. Further details on this procedure are given in Appendix \ref{sec:tests_inversion}. 

\begin{figure*}
    \centering
    \sidecaption
    \includegraphics[width=12.5cm]{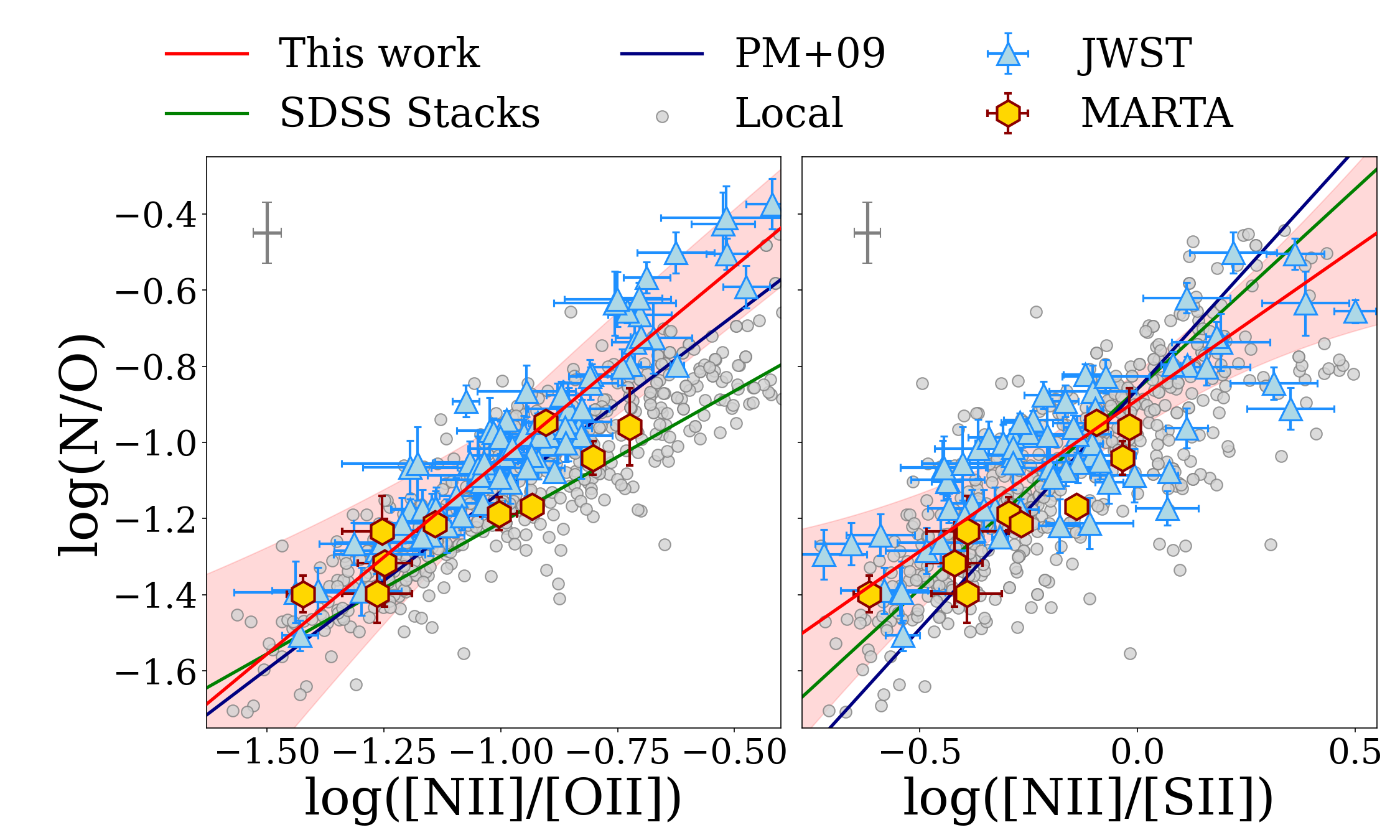}
    \caption{Relation between the N/O ratio measured using the $T_e$ method and the N2O2 (upper panel) and N2S2 (lower panel) diagnostics for our full high-$z$ auroral-line sample. For comparison, we include local galaxies with auroral-line detections from the literature as gray points (see references in Sec. \ref{sec:local_objects}). 
    The local calibration from \citet{perez-montero_impact_2009} is shown in blue and a fit to the SDSS stacks of \cite{curti_new_2017} in green. In both panels, the grey error bar in the top-left corner represents the median measurement uncertainty of the local comparison sample. The red solid line shows the best fit obtained from the high-z sample with the uncertainty range shown as a shading.}

    \label{fig:N2O2_N2S2}
\end{figure*}

\subsection{Nitrogen abundances from direct measurements}
\label{sec:N_determination_direct}

For the galaxies with a direct electron temperature measurement, nitrogen abundances were computed from the dust-corrected fluxes of the \NII~$\lambda6584$ and \OII~$\lambda\lambda3727,3729$ emission lines. We assumed that the ionic abundances of N$^+$ and O$^+$ trace the nitrogen and oxygen content in the low-ionisation zone of the nebula, and adopted the same electron temperature ($T_2$) for both ions. This choice is justified by the similar ionisation potentials required both to form these ions from their neutral states (14.5 eV for N$^0 \rightarrow$ N$^+$ and 13.6 eV for O$^0 \rightarrow$ O$^+$) and to ionise them further (29.6 eV for N$^+ \rightarrow$ N$^{2+}$ and 35.1 eV for O$^+ \rightarrow$ O$^{2+}$), ensuring they originate from roughly the same physical region \citep[e.g.,][]{garnett_nitrogen_1990, vila_costas_nitrogen-ratio_1993}.

As detailed in Sec. \ref{sec:met_determination_direct} we measure $T_2$ directly when possible. This was feasible for only a small subset of the high-redshift sample (9 objects), and for $\sim$900 local Universe galaxies. Alternatively we infer it from the $T_2$–$T_3$ relation of \citetalias{Cataldi2025}.
We also tested how the results would change when using $T_e$(\NII) estimated from the empirical relation between $T_e$(\NII) and $T_e$(\OIIIopt) available in the literature. This relation is well-calibrated and relatively tight in the low-redshift Universe (no auroral lines from \NII are available for the high-$z$ sample), as shown in previous studies \citep[e.g.,][]{garnett_electron_1992, perez-montero_impact_2009}. When adopting $T_e$(\NII) for N$^+$ and $T_e$(\OII) for O$^+$ the derived log(N/O) increases by $\sim$ 0.06 dex.

We computed the emissivities of the relevant lines using the adopted $T_e$ and $n_e$.
To obtain the total N/O abundance ratio, we corrected for unseen ionisation states by applying an ionisation correction factor (ICF). We adopted the ICF prescription from \citet{amayo_ICFs_2020}, which accounts for the relative ionisation structure of nitrogen and oxygen based on photoionisation model grids and is expressed as a function of the nebular excitation $\omega$, defined as $\omega = \mathrm{O}^{++}/(\mathrm{O}^+ + \mathrm{O}^{++})$.

While it is true that such ICF prescriptions could evolve with cosmic time, we have tested whether their application introduces artificial trends in our results by repeating all analyses without applying any ionisation correction. This latter approach is equivalent to assuming N$^+$/O$^+$ = N/O, which is the ICF prescription proposed by \citet{peimbert_chemical_1969} based on the similar ionisation energies of N and O, and is commonly adopted in the literature when deriving N/O from singly ionised species. The ICF values in our sample are typically modest, up to 1.2 with a median value of $\sim$1.1. We find that removing the ICF corrections does not alter any of the conclusions presented in this work, introducing at most a small systematic offset (0.07 dex higher) in the absolute N/O values. This demonstrates that our results are not affected by artifacts introduced by potential ICF redshift evolution.

The total N/O abundance ratio was then computed as:
\begin{equation}
\log(\mathrm{N/O}) = \log\left( \frac{F_{\mathrm{[N\,II]}\,\lambda6584}}{F_{\mathrm{[O\,II]}\,\lambda\lambda3727,3729}} \frac{\varepsilon_{\mathrm{[O\,II]}\,\lambda\lambda3727,3729}\,(T_e, n_e)}{\varepsilon_{\mathrm{[N\,II]}\,\lambda6584}\,(T_e, n_e)}  \mathrm{ICF} \right),
\end{equation}
where $F_{\mathrm{[N\,II]}\,\lambda6584}$ and $F_{\mathrm{[O\,II]}\,\lambda\lambda3727,3729}$ are the reddening-corrected fluxes of the \NII and \OII lines; $\varepsilon_{\mathrm{line}}(T_e, n_e)$ denotes the emissivity of each line at a given electron temperature and density (computed with \texttt{getEmissivity} function from \texttt{PyNeb}), and $\mathrm{ICF}$ is the ionisation correction factor.

\section{Results}
\label{sec:results}

\subsection{Strong-line diagnostics for N/O at high redshift}\label{sec:N2O2_calib}

The direct derivation of nitrogen abundances relies on the measurement of electron temperatures and, thus, on the detection of faint auroral lines. However, due to the intrinsic weakness of these features and observational limitations - especially at high redshift - such measurements are often unfeasible. For this reason, several studies have proposed empirical calibrations that relate strong-line ratios to the N/O ratio, allowing for indirect determinations when auroral lines are not available.

Among the most widely adopted N/O indicators is the N2O2 ratio.
This diagnostic is physically motivated by the similar ionisation potentials of N and O, as mentioned in Section \ref{sec:N_determination_direct}.
Its reliability and extensive use in the literature have made it the de facto standard for estimating N/O when direct methods are not feasible \citep[e.g.,][]{van_zee_spectroscopy_1998, perez-montero_impact_2009, hayden-pawson_NO_2022}.

A complementary diagnostic is N2S2, which has been widely adopted as a tracer of N/O in numerous studies \citep[e.g.,][]{perez-montero_impact_2009,2018ApJ...868..117S, hayden-pawson_NO_2022}. N2S2 has the advantage of being largely insensitive to dust attenuation, owing to the close proximity in wavelength of the two lines involved. Its physical interpretation is, however, less straightforward, as it is affected by differences in the ionisation structure of N and S and variations in the S/O ratio. Nonetheless, it remains a useful proxy for N/O.

Figure \ref{fig:N2O2_N2S2} shows the relation between the directly measured N/O ratio and the N2O2 (left) and N2S2 (right) diagnostics for our full auroral-line dataset, as described in Section \ref{sec:auroral_data}. 
We also compare our results with the local calibration from \citet{perez-montero_impact_2009} and with a newly derived relation based on SDSS stacked spectra from \citet{curti_new_2017}. This SDSS calibration was obtained following the work of \citet{hayden-pawson_NO_2022}, but re-deriving the abundances using the same ICFs and atomic data consistently adopted throughout our analysis. Comparing the original \citet{hayden-pawson_NO_2022} calibration directly with our dataset would introduce an apparent offset, primarily driven by differences in the adopted ICFs, as no ionisation corrections were applied in their work. When instead recalibrated on the same SDSS stacks with our consistent assumptions, the resulting relation shows consistency with our full dataset. 
The bestfit relation for the SDSS sample is then

\begin{equation}
    \log(\mathrm{N/O}) = (0.60 \pm 0.07) \times \mathrm{N2O2} - (0.63 \pm 0.09).
    \label{eq:n2o2_SDSS}
\end{equation}

As shown in Figure~\ref{fig:N2O2_N2S2}, the high-redshift galaxies in our sample lie within the locus defined by local sources and are broadly consistent with the trend from previous empirical calibrations. However, while they formally fall within the scatter of the local distribution, in the N2O2 diagram, high-z objects predominantly occupy the upper envelope of the distribution, showing systematically elevated N/O values at fixed N2O2 and hinting at a possible redshift evolution of this relation. Conversely, in the N2S2 diagram they appear more scattered throughout the local distribution and exhibit better overall consistency with nearby galaxies.

To quantify their behaviour, we performed a linear regression for both the diagnostics. The regressions were performed on the whole $z>1$ sample. For consistency, we verified that splitting the sample in two redshift ranges ($1<z<4$; $4<z<6$) gives comparable coefficients within their uncertainties (see also \hyperref[sec:scatter_NO_strong_lines]{Appendix B} for further analysis).
The best-fit relations were obtained using a Huber regressor, a robust algorithm that combines the properties of ordinary least-squares and absolute-deviation fitting. It behaves like least-squares for well-behaved data points but progressively downweights outliers beyond a certain threshold, providing stable estimates even in the presence of intrinsic scatter or uncertain measurements-conditions typical of high-redshift spectroscopic observations.
Uncertainties on the slope and intercept were estimated via the jackknife resampling technique, which recomputes the fit after iteratively removing one data point at a time. This method is preferred over bootstrap resampling given the relatively small size of the high-z sample, which would otherwise lead to excessive repetition of individual objects across subsamples.

Our fiducial best-fit relation 
is then
\begin{equation}
\log(\mathrm{N/O}) = (1.02 \pm 0.09) \times \mathrm{N2O2} - (0.03 \pm 0.06).
\label{eq:n2o2_calibration}
\end{equation}
The estimated intrinsic scatter of this relation is $\sigma_{\mathrm{int}} = 0.07$~dex.
The resulting best-fit relation is therefore slightly steeper than that derived from SDSS stacks, reflecting the tendency of our high-redshift galaxies to lie along the upper envelope of the local distribution. Nevertheless, the coefficients remains consistent within the uncertainties with the local calibration of \citet{perez-montero_impact_2009}.

A similar conclusion holds for the N2S2–N/O relation (right panel of Figure~\ref{fig:N2O2_N2S2}), despite its slightly larger intrinsic scatter of 0.10 dex compared to N2O2: the high-redshift galaxies -- both from the MARTA sample and from the literature -- are consistent with the distribution of local galaxies and the calibrations derived in the nearby Universe. The newly derived empirical fit are in good agreement with local trends, supporting the use of N2S2 as a secondary tracer of N/O in distant galaxies. 
The best-fit N2S2–N/O relation is 
\begin{equation}
\label{eq:n2s2_calibration}
\log(\mathrm{N/O}) = (0.80 \pm 0.16) \times \mathrm{N2S2} - (0.89 \pm 0.08).
\end{equation}

Overall, these results indicate that the empirical N2O2 and N2S2 relations therefore provide reliable diagnostics for estimating N/O in galaxies out to the epoch of reionisation. 
The origin of the scatter observed in the N/O-N2O2 and N/O-N2S2 strong-line diagnostics is further investigated in \hyperref[sec:scatter_NO_strong_lines]{Appendix B}.

\begin{figure*}
    \centering
    \includegraphics[width=0.95\linewidth]{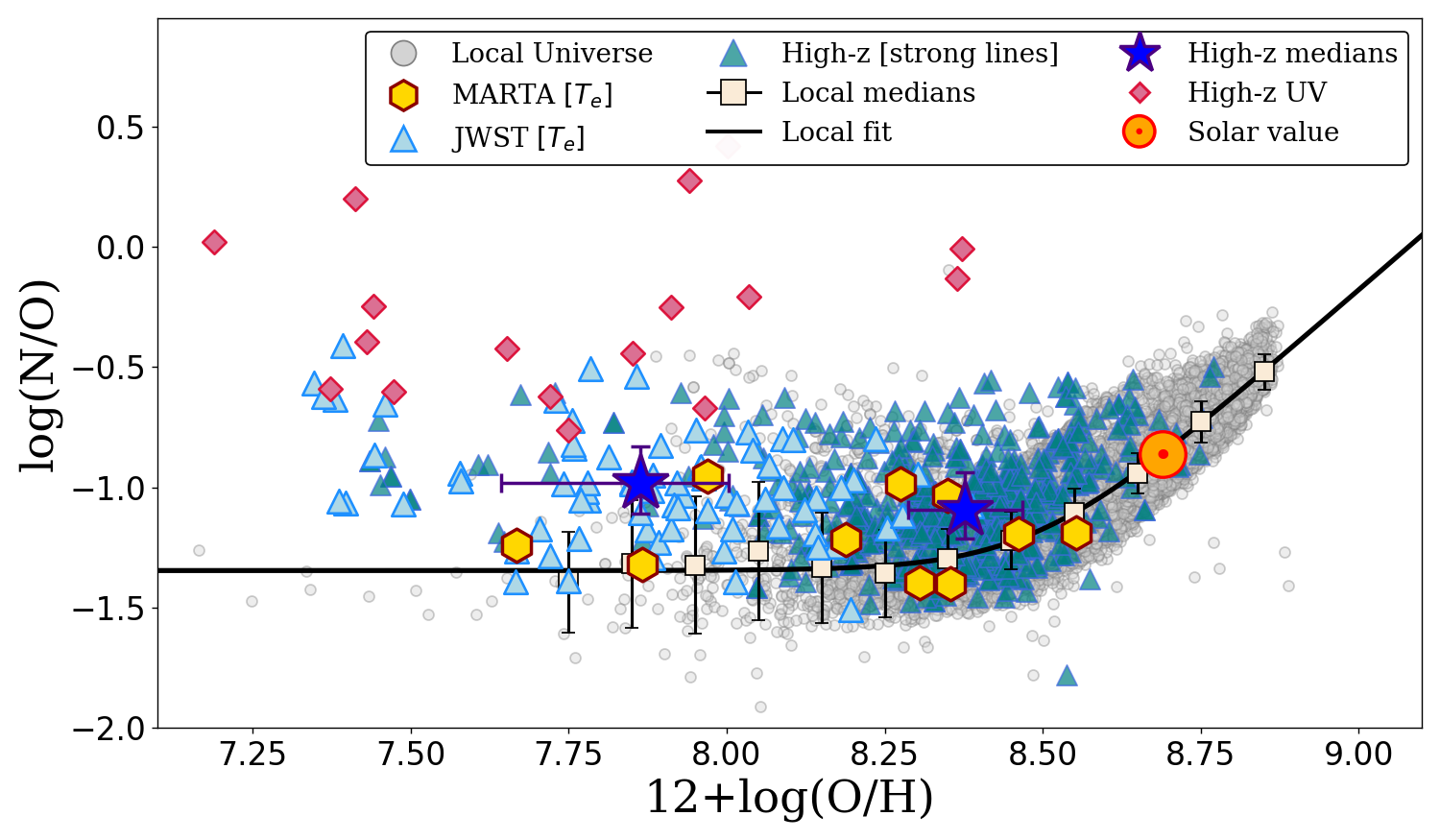}
    
    \caption{Relation between N/O and O/H for our full sample, as described in Section~\ref{sec:data}. Gold hexagons indicate MARTA galaxies with direct measurements, and light-blue triangles show the JWST auroral-line dataset.  High-redshift strong-line galaxies from the literature are plotted as teal triangles, and local galaxies (both direct and strong-line measurements) as grey circles. White squares represent local galaxies binned in metallicity, showing the median N/O value in each bin. Two large blue stars indicate the median N/O values for high-redshift galaxies, with the first corresponding to $12+\log(\mathrm{O/H}) < 8.1$ and the second to $12+\log(\mathrm{O/H}) \ge 8.1$. The black line shows the best-fit relation derived from the binned values of local galaxies using the functional form from \citet{hayden-pawson_NO_2022}, with the best-fit parameters obtained in this work and reported in Table~\ref{table:fit}. The solar abundance values from \citet{Asplund21} are shown with the orange Sun symbol. Finally, the pink diamonds represent higher-redshift, UV-based abundances drawn from the recent compilation by \citet{Ji2026}. These sources are shown to illustrate the respective loci occupied by optical-based and UV-based abundances in this parameter space.
    }

    \label{fig:NO_metallicity_bestfit}
\end{figure*}

\begin{SCfigure*}
    \centering
    \includegraphics[width=1.35\linewidth]{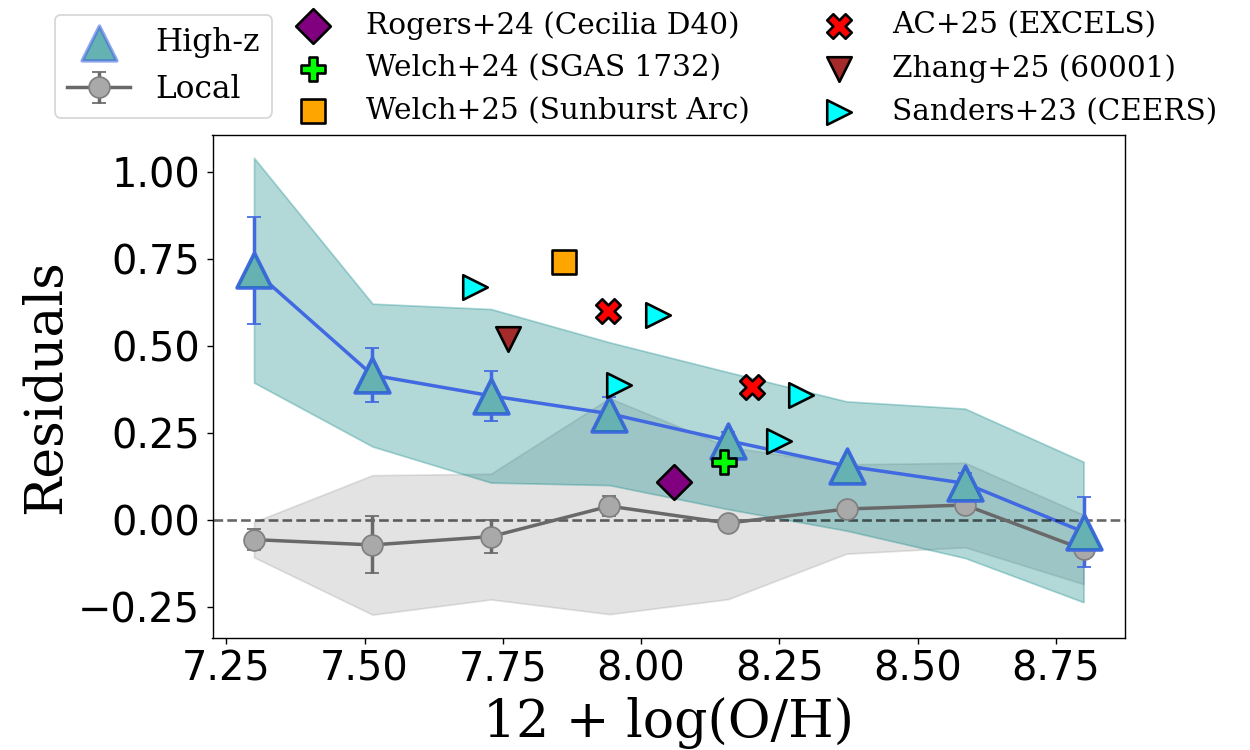}
    \caption{Residuals of the observed N/O ratios with respect to the local best-fit relation, binned as a function of metallicity for the local (grey) and high-redshift (teal) samples. Error bars represent the uncertainties on the median values, while the shaded regions indicate the dispersion within each bin. The single data points represent the residuals of previously published literature sources with N/O and O/H measurements from rest-frame optical lines: five objects from \citet{sanders_ceers_2023} (CEERS program, cyan triangles); two galaxies from the EXCELS programme in \citet{2025MNRAS.540.2991A} (red crosses); the Sunburst Arc from \citet{2025ApJ...980...33W} (orange square); the SGAS~1732 from \citet{2024ApJ...975..196W} (green plus); the GTO\_2758\_60001 from \citet{Zhang2026} (dark-red inverted triangle); and the CECILIA~D40 from \citet{2024ApJ...964L..12R} (purple diamond).}

    \label{fig:NO_metallicity_residuals}
\end{SCfigure*}

\subsection{N/O-O/H relationship}
\label{sec:no_vs_oh}

In this section we investigate the relation between the N/O ratio and oxygen abundance for our sample of high-redshift systems, and include a direct comparison with the trend followed by low-redshift galaxies. In particular, we build upon the direct-method measurements presented in the previous section, and expand our analysis by including the strong-line datasets described in Sections~\ref{sec:strong_line_data}. For these galaxies, as illustrated in Figure~\ref{fig:scheme}, we apply the methodology outlined in Section~\ref{sec:oxygen_from_strong_lines} to derive oxygen abundances and applied the newly derived calibration based on N2O2 given by Equation \ref{eq:n2o2_calibration} for the N/O abundances.
This approach allows us to combine direct auroral-line measurements with strong-line diagnostics calibrated against the direct method, ensuring internal consistency across the entire sample.

In Figure~\ref{fig:NO_metallicity_bestfit} we plot log(N/O) versus $12+\log(\mathrm{O/H})$ for both our high-$z$ sample and local reference galaxies. MARTA galaxies with auroral-line detections are highlighted as gold hexagons, compiled JWST galaxies with T$_{\text{e}}$-based measurements are shown as light-blue triangles, and high-$z$ objects with strong-line-based abundance estimates are marked as teal triangles. For reference, we also show higher-redshift galaxies drawn from the recent compilation by \citet{Ji2026} (pink diamonds), for which N/O estimates are based on UV nitrogen lines. These values are taken directly from the literature, without any re-derivation, and are mainly included to illustrate the respective loci occupied by optical-based and UV-based abundances in this parameter space. 
The local reference sample, comprising both direct and strong-line measurements, is represented by gray circles. The white squares show median values of the local sample binned in metallicity, while the large blue stars indicate the median values of the high-$z$ sample divided into two metallicity regimes ($12+\log(\mathrm{O/H}) < 8.0$ and $\geq 8.0$). The solid black curve represents the best-fit N/O--O/H relation to the median-binned local dataset, adopting the following functional form:
\begin{equation}
\label{eq:fit}
\log(\mathrm{N/O}) = \log(\mathrm{N/O})_0 + \frac{\gamma}{\beta} \, 
\log \left[ 1 + \left( \frac{(\mathrm{O/H})}{(\mathrm{O/H})_0} \right)^\beta \right] \, .
\end{equation}

This parametrization was introduced by \cite{hayden-pawson_NO_2022} to describe the log(N/O) versus $12+\log(\mathrm{O/H})$ relation. It provides a smooth transition from the N/O plateau at low metallicity (assumed to trace primary nitrogen enrichment) to a power law behaviour for $ \mathrm{O/H} > \mathrm{O/H}_0$ (corresponding to the regime where secondary nitrogen enrichment dominates). The power law component has slope $\gamma$, and the sharpness of the transition between the two regimes is regulated by the parameter $\beta$. The best-fit values of the fit parameters are presented in Tab. \ref{table:fit}.

\begin{table}[t!]
\centering
\caption{Best-fit parameters of the N/O–O/H relation for the local galaxy sample.}

\label{table:fit}
\begin{tabular}{l c c c c}
\hline
$\log(\mathrm{N/O})_0$ & $12+\log(\mathrm{O/H})_0$ &  $\gamma$ & $\beta$ \\
\hline
$-1.385 \pm 0.017$ & $8.43 \pm 0.08$  & $1.8 \pm 0.1$ & $5.6 \pm 0.9$ \\
\hline
\end{tabular}
\tablefoot{
The coefficients refer to equation~\ref{eq:fit}. $12+\log(\mathrm{O/H})_0 = 8.43$ corresponds to $(\mathrm{O/H})_0 = 2.7 \times 10^{-4}$.}
\end{table}

Figure~\ref{fig:NO_metallicity_bestfit} demonstrates a systematic nitrogen enhancement of the high-$z$ galaxy population relative to the local relation, particularly at lower metallicities. This is highlighted by the median N/O points as measured in the two different O/H regimes, since both lie above the local fit. 

To quantitatively assess the difference between the two distributions, we compared the residuals of the high-$z$ and local samples relative to our best-fit $z=0$ relation (see Appendix~\ref{sec:statistics}). The two distributions are statistically inconsistent with being drawn from the same parent population: the Anderson-Darling test\footnote{The Anderson–Darling test is a non parametric test that, similarly to the Kolmogorov–Smirnov, compares the cumulative distributions of two samples. Unlike K-S, it assigns more weight to differences in the tails of the distributions and provides a more uniform sensitivity across quantiles. This makes it particularly suitable when the samples have very different sizes or when discrepancies in the extreme values are scientifically relevant.} \citep[AD-test,][]{scholtz&stevens1987} rejects the null hypothesis at high significance ($p < 10^{-3}$). Additional tests and a full discussion are presented in Appendix~\ref{sec:statistics}.

In Figure~\ref{fig:NO_metallicity_residuals} we show the residuals in $\log(\mathrm{N/O})$ from the best-fit model as a function of $12+\log(\mathrm{O/H})$, binning the data in $\approx0.2$~dex intervals and computing median residuals separately for the local and high-$z$ samples. A clear divergence emerges between the two populations: high-redshift galaxies exhibit increasingly positive residuals toward lower metallicities, indicating that the enhancement in N/O at high redshift is most pronounced in metal-poor systems. This behaviour becomes particularly evident at $12+\log(\mathrm{O/H}) \lesssim 8.0$, where the median high-$z$ residual reaches $\sim0.3$~dex.

To facilitate a more direct comparison, we also include previously published abundance measurements at these redshift based on rest-frame optical lines. These reference points are drawn from the CEERS program, with emission-line fluxes taken from \citet{sanders_ceers_2023}; the EXCELS program \citep{2025MNRAS.540.2991A}; the Sunburst Arc from \citet{2025ApJ...980...33W}; SGAS~1732 from \citet{2024ApJ...975..196W}; the GTO\_2758\_60001 source from \citet{Zhang2026}; and the CECILIA~D40 galaxy from \citet{2024ApJ...964L..12R}. For each of these objects, we re-derived the abundances from their published emission-line fluxes, following the methodology described in Section~\ref{sec:ISM_prop_sec_3}. The recomputed metallicities and N/O ratios show good agreement with the values reported in the original works, typically within the uncertainties. 

We mostly confirm conclusions from previous works regarding nitrogen enrichment. Systems independently reported in the literature as nitrogen enhanced (e.g. the Sunburst arc, the \citet{Zhang2026} source, EXCELS objects) are also found to lie above the local trend in our analysis, and in some cases even lie outside the dispersion of our high-redshift trend. Galaxies reported to follow the local pattern (CECILIA~D40, SGAS~1732) fall within the dispersion of the local population. 

\cite{2025ApJ...981..136S} classify only three of the CEERS sources as N-enhanced, noting that the remaining two are consistent with the dispersion of the local relation. With our larger statistics we find all CEERS sources to be N-enhanced with respect to the local sample, and two of them to lie above the dispersion of the high-redshift one. Our findings therefore provide context for previous discussions of individual sources in the literature, confirming the existence of both N-enhanced and non-enhanced sources, and showing how their frequency changes with metallicity. 

\begin{SCfigure*}
    \centering
    \includegraphics[width=1.3\linewidth]{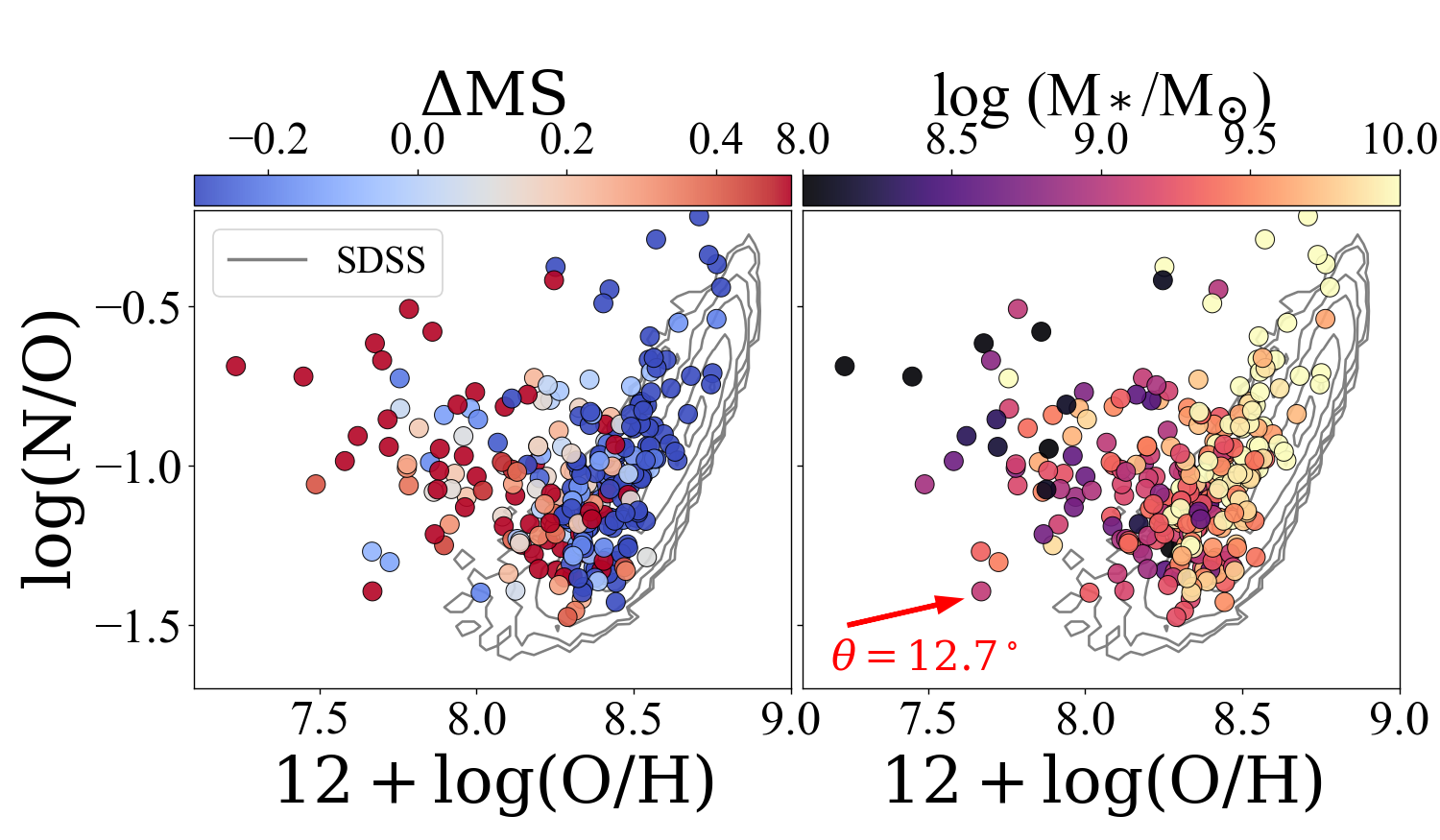}
    \hspace{0.2em}
    \caption{Same as Figure \ref{fig:NO_metallicity_bestfit}, but with the SDSS density contours as the sole local reference for simplicity. The high-$z$ sample is represented with circles, with no distinction between different datasets. Left panel: the high-$z$ sample is colour-coded by $\Delta \text{MS}$, i.e. the deviation from the parametrised SFMS from \citet{popesso_SFMS_2023}, computed at each specific redshift. 
    Right panel: the high-$z$ sample is colour-coded by stellar mass. The gradient angle of the $M_*$ trend within the plane is shown as a red arrow, visually representing the finding that N-enhanced galaxies are preferentially associated with lower stellar masses.
    }

    \label{fig:NO_OH_color_coding}
\end{SCfigure*}

\section{Discussion}
\label{sec:discussion}

While our findings are certainly affected by observational biases and selection effects (discussed in Appendix \ref{sec:N_enhanced}), we are in agreement with recent studies pointing to moderate nitrogen over-enhancement for the bulk galaxy population at high $z$ \citep{masters_physical_2014, strom_nebular_2017, hayden-pawson_NO_2022, 2025ApJ...982...14H, Cameron_26}. However, those results were based on strong-line measurements or stacks, and thus the present work -- incorporating direct-method abundances from auroral lines from individual galaxies, alongside consistently calibrated strong-line diagnostics -- provides a more robust confirmation of the underlying trend.

Various scenarios have been proposed in the literature to explain elevated N/O ratios at low metallicity. For example, bursty star formation, where the temporal decoupling between oxygen production (dominated by massive stars on few Myr timescales) and nitrogen release (from intermediate-mass stars on $\sim$100 Myr timescales) can lead to transiently high N/O ratios \citep{2005A&A...434..531K, belfiore_p-manga_2015,  Bhattacharya26, McClymont2026}. If star formation proceeds in short, intense bursts followed by quiescent phases, oxygen may be injected early while nitrogen continues to enrich the ISM after the burst has faded, temporarily boosting N/O. 
In this scenario, as pointed out by \cite{McClymont2026}, N/O overabundance should correlate with a negative offset with respect to the MS.

In the left panel of Figure \ref{fig:NO_OH_color_coding}, we show the N/O-O/H relation colour-coded by $\Delta MS$, which represents the offset of each galaxy SFR from the star formation main sequence (MS) prediction. Whenever stellar masses were provided in the original publications, we adopted those values; otherwise, we used the estimates from the DJA catalogue, obtained from SED fitting as described in \cite{Valentino23} using \textsc{eazy-py} \citep{Brammer_EAZY_2008}. SFRs were estimated then from the \Halpha\ luminosity using the conversion of \citet{kennicutt_star_1998}, scaled to a Chabrier initial mass function \citep{chabrier_galactic_2003}.
We adopt the MS relation from \citet{popesso_SFMS_2023}, evaluating it at the redshift of each individual galaxy in order to compute the corresponding $\Delta MS$. 
A high fraction of N-enhanced objects in our sample exhibit positive $\Delta MS$, rather than the suppressed star formation activity expected if the N/O enhancement were driven by a temporary quenching phase following a starburst. This trend may, however, be biased by observational selection effects, so we caution against over-interpreting it at this stage. It is also worth noting that, although the majority of galaxies ($\gtrsim$50\%) with enhanced N/O lie above the main sequence, a non-negligible fraction is found below it, and overall we do not observe any clear correlation between $\Delta \log(\mathrm{N/O})$ and $\Delta MS$, as discussed in greater detail in Appendix \ref{sec:N_enhanced}.

A related scenario involves the selective removal of metals by Type II supernova-driven outflows. Previous studies \citep[e.g.][]{2024ApJ...970...14S, 2025A&A...699A...6P} suggest that energetic supernova feedback can eject oxygen-rich material more efficiently than the nitrogen released later by LIMS. Simple chemical evolution models show that such differential outflows can explain N over-enhancement \citep{2025A&A...697A..96R}. \cite{McClymont2026} argue that bursty star formation naturally produced differential outflows when integrated over time, because the strongest feedback phase acts on the ISM after oxygen is newly synthetized. The gas-poor state following the burst can then be enriched in nitrogen by LIMS.

A second class of explanations involves changes in gas metallicity driven by the inflow of pristine material. The accretion of metal-poor gas dilutes the ISM oxygen abundance, while leaving N/O essentially unaffected. This mechanism is expected to be particularly relevant at high redshift, when cosmological accretion rates are higher and cold flows are predicted to be common \citep{dekel_cold_2009, 2010Natur.463..781T, 2012MNRAS.421...98D}. In this picture, galaxies experiencing substantial inflows should exhibit lower O/H than typical systems of the same stellar mass, while their N/O ratios should reflect the pre-inflow chemical state. In particular, if the pre-inflow metallicity lies in the regime where secondary nitrogen production dominates, a galaxy can retain a high N/O ratio even after its O/H is temporarily diluted by a massive inflow of metal-poor gas.

In the right panel of Figure \ref{fig:NO_OH_color_coding} we examined the N/O–O/H relation colour-coded by stellar mass. 
If dilution by pristine gas were the dominant mechanism, the most N-enhanced systems should also be, generally, more massive: a pure inflow event primarily reduces O/H, shifting galaxies horizontally toward lower metallicity at fixed mass. In this scenario, a galaxy would retain the mass and N/O expected for an evolved, enriched system, while appearing more metal-poor. Conversely, we observe that lower-mass galaxies tend to have lower metallicities and vice versa, broadly consistent with the expectations from the mass-metallicity relation, without obvious systematic deviations that would suggest pristine gas dilution as the primary driver.
We quantified the stellar mass dependence across the N/O-O/H plane by calculating the gradient angle\footnote{The gradient angle and significance were derived by fitting the directional derivative of M in the 2D plane and estimating uncertainties through bootstrap reshuffling of the mass values.}, obtaining a value of $12.7^{\circ} \pm 6.3^\circ$. The associated statistical significance ($p_{\text{value}} \lesssim 10^{-6}$) confirms the robustness of this trend.
This trend confirms that N-enhanced galaxies tend to be preferentially associated with lower stellar masses, which runs opposite to the expectations under an idealized pristine gas dilution scenario. 
The current statistics are limited, and a larger sample will be needed to confirm this behaviour robustly. Nevertheless, the observed trend disfavours pristine-gas inflow as the primary driver of the nitrogen enhancement \citep[see also][]{hayden-pawson_NO_2022}. 

Another relevant class of models invokes intermittent, non-monotonic star-formation histories coupled with galactic winds. For example, a dual-burst implementation was developed by \citet{kobayashi_ferrara_gnz11_2024} to explain the extremely N-enhanced system GN-z11 at $z=10.6$ \citep{bunker_gnz11_2023, cameron_gnz11_2023}, where early nitrogen excess is attributed to Wolf–Rayet (WR) enrichment \citep[see also][for an application of such a model to a $z=9.4$ galaxy]{Curti_gsz9_25}. Furthermore, direct observational signatures of WR stars -- such as broad blue and orange emission bumps -- have also been detected in several N-enhanced systems \citep[e.g.,][]{Rivera-Thorsen2024, Berg2025, Morishita2025}.
However, such a temporary nitrogen enhancement scenario requires specific timing: WR stars, which are massive stars ($\gtrsim 25 M_{\odot}$) in a very short-lived evolutionary phase, emerge only within a few Myr after a starburst and enrich the ISM on similarly rapid timescales (e.g., \citealt{Schaerer1998, Crowther2007}). Although such coincidences may occur in individual systems, it is unlikely that this mechanism could operate in a coordinated way across a large galaxy sample, making it implausible as the primary driver of the systematic N/O elevation we observe.

A similar dual-burst framework was applied by \citet{Bhattacharya26} to reproduce the abundance patterns of low-redshift (\,<\,$z$\,>\,=\,0.1), nitrogen-enriched, low-metallicity galaxies observed in DESI.
We note that, contrary to \citet{kobayashi_ferrara_gnz11_2024}, in this case the observed N/O enhancement is a consequence of the combined effect of delayed N-enrichment from AGB stars and strong outflows occurring after the second burst of star-formation (which follows a gas accretion episode that significantly diluted the metallicity of the system), while the fast N-enhancement from WR stars\footnote{which happens before the O production from CCSNe}, although included in the model, does not produce significant observable features due to the very short timescales involved.

\begin{figure}
    \centering
    \includegraphics[width=\linewidth]{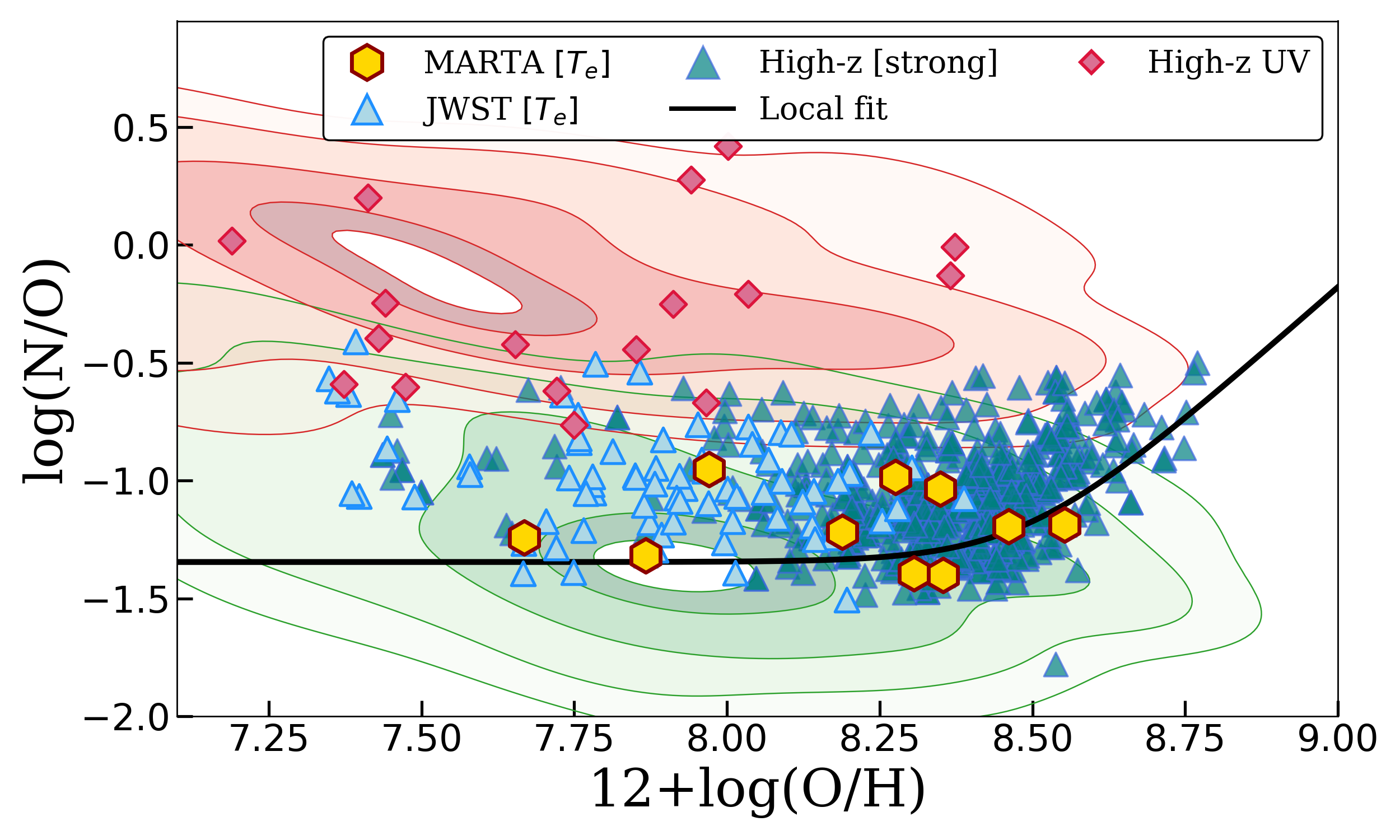}
    
    \caption{Same as Figure \ref{fig:NO_metallicity_bestfit}, but with Milky Way globular cluster contours (representing probability levels of 5, 16, 50, 84, and 95 percent), respectively of the first (in green) and second (in red) generation of stars. The contours are taken from \citet{Ji2026}, extracted from the APOGEE catalogue in SDSS DR17. 
    }

    \label{fig:NO_metallicity_contours}
\end{figure} 

Finally, several studies have proposed a link between nitrogen-enhanced high-redshift galaxies and the progenitors of present-day globular clusters (GCs) \citep[e.g.,][]{ charbonnel_gnz11_2023, Senchyna2024, 2025A&A...700A.265D}. The unusual abundance patterns in some early galaxies resemble those in Milky Way GCs, suggesting that similarly dense, compact stellar systems might have influenced their chemical evolution. This connection was strengthened by \citet{Ji2026}, who showed that the most N-enhanced $z>4$ galaxies whose abundances are measured with UV nitrogen lines share chemical signatures with second-generation (2G) GC stars.
As shown in Figure \ref{fig:NO_metallicity_contours}, our comparison with Galactic GC populations reveals that our high-z ($z\sim2$–4) sample, particularly direct-measurement galaxies, overlaps better with first-generation (1G) GC stars in the N/O–O/H plane, contrasting with the most metal-poor, ultra–N-enhanced $z>4$ systems, which overlap with 2G stars \citep{Ji2026}, highlighting that our objects display only moderate enhancement, incompatible with the extreme 2G-like signatures in the most UV-bright high-$z$ outliers. We note, however, that the median properties of 1G GC stars are broadly consistent with the local N/O-O/H relation. Our galaxies therefore fall within the intrinsic scatter of the 1G distribution rather than tracing its median trend. A simple quantitative test shows that $\sim$70\% of the high-z objects lie within the 84th percentile of the 1G distribution, although their median properties remain offset.

In this context, recent work suggests that 1G GC star formation may have occurred later than 2G \citep{2024ApJ...977...14C, 10.1093/mnras/staf745}, making a resemblance between 1G-like abundances and lower-redshift galaxies plausible. The physical origin of extreme nitrogen enrichment in 2G stars remains debated \citep[e.g.,][]{charbonnel_gnz11_2023}, with scenarios such as super-massive star contributions still under discussion. This comparison should thus be regarded as phenomenological rather than evidence for a specific mechanism.

\section{Conclusions}
\label{sec:conclusions}

We presented a comprehensive investigation of the nitrogen enrichment in $\sim 500$ star-forming galaxies at redshift $z \approx 2-4$, combining ultra-deep JWST/NIRSpec spectroscopy from the MARTA survey with a large compilation of both high-$z$ and local Universe literature data. 
By directly detecting faint auroral lines in galaxies, we determined gas-phase abundances through the direct electron-temperature method, and recalibrated empirical strong-line relations for N/O at high redshift for the first time. 
Our work thus provides a homogeneous framework for comparing nitrogen enrichment across cosmic time, bridging the local Universe and galaxies up to $z \sim 8$. In particular, our main results include:

\noindent\emph{Sample and methodology}.  The analysis combines 76 high-$z$ galaxies with auroral-line detections and 430 additional high-$z$ systems with strong-line measurements, together with $\sim9\times10^4$ local galaxies for comparison (Sect.\ref{sec:data}; Figure~\ref{fig:z_hist}). Direct T$_e$-based abundances were derived when possible, while strong-line calibrations were applied elsewhere, following the self-consistent framework summarised in Figure~\ref{fig:scheme}.\\
\emph{Calibration of N/O strong-line diagnostics} (Sect. \ref{sec:N2O2_calib}; Figure~\ref{fig:N2O2_N2S2}). We recalibrated the N2O2 and N2S2 diagnostics at high redshift using galaxies with direct N/O measurements.  
No systematic offset is observed relative to local calibrations, confirming that strong-line N/O indicators remain valid up to $z \sim 6$.  Among the two diagnostics, N2O2 provides the tightest correlation, with an intrinsic scatter of $\sigma_{\rm int} \approx 0.09$~dex, while N2S2 is slightly less tight ($\sigma_{\rm int} \approx 0.13$~dex) but has the advantage of being less sensitive to dust extinction effects.\\
\emph{The N/O-O/H relation across cosmic time} (Sect. \ref{sec:no_vs_oh}; Figure~\ref{fig:NO_metallicity_bestfit}, \ref{fig:NO_metallicity_residuals}). Combining all datasets, we constructed the most extensive N/O-O/H relation for galaxies at $z>1$; we fitted the functional form from \citet{hayden-pawson_NO_2022} (Equation \ref{eq:fit}) to the local Universe dataset, finding the best-fit parameters reported in Table \ref{table:fit}.  High-redshift galaxies display a systematic enhancement of N/O relative to the local relation, by $\Delta(N/O) \approx 0.18$~dex (median), confirmed at $>99.9$\% confidence (AD and Mann--Whitney tests). The enhancement is metallicity-dependent, reaching $\sim$0.3--0.4~dex at $12 + \log(\mathrm{O/H}) \lesssim 8.1$ (Figure~\ref{fig:NO_metallicity_residuals}). \\
\emph{Physical interpretation of N/O enhancement} (Sect. \ref{sec:discussion}; Figure \ref{fig:NO_OH_color_coding}, \ref{fig:NO_metallicity_contours}). We explored several physical mechanisms proposed to explain the elevated N/O ratios at high redshift. Most N-enhanced galaxies in our sample lie above the main sequence (Figure \ref{fig:NO_OH_color_coding}, right panel), which appears inconsistent with the expectation that bursty star formation followed by quiescent phases drives the N/O enhancement, since such scenario would predict suppressed star formation activity. We also find that N-enhanced systems are preferentially associated with lower stellar masses (Figure \ref{fig:NO_OH_color_coding}, left panel), disfavouring pristine gas dilution as the main driver of the enhancement. 
Finally, comparison with Galactic globular cluster stellar populations (Figure \ref{fig:NO_metallicity_contours}) shows that our $z\sim2$–4 systems lie within the scatter of first-generation GC stars, yet their median N/O at a given metallicity exceeds the 1G median trend, and remain well separated from the locus of 2G-like, ultra–N-enhanced high-redshift galaxies. This indicates that the moderate N/O excess we observe is consistent with relatively mild processing in compact star-forming regions, without invoking the extreme enrichment associated with the most chemically exceptional early-universe systems.

Overall, this study establishes a comprehensive view of nitrogen enrichment at Cosmic Noon ($<z>= 3.0$), revealing the the overall population is mildly nitrogen enhanced, especially at low metallicity. Larger samples are needed to reduce observational bias, which may be driving some of the population properties observed here. However, taken at face value, none of the scenarios outlined in the literature fit our data in detail, outlining the need for renewed theoretical efforts in the field of chemical evolution modelling at Cosmic Noon and not only at $z>4$. The measurement of abundances of additional species will provide even more stringent constrains on theoretical work and greatly contribute to our understanding of how star formation proceeded from Cosmic Dawn to Noon.

\begin{acknowledgements}
FB, FM, AM, MG, IL, CM, AF, EB, GC acknowledge support from the INAF Fundamental Astrophysics programme 2022, 2023 and 2024.
IL, FB, and AM acknowledge support from PRIN-MUR project “PROMETEUS”  financed by the European Union -  Next Generation EU, Mission 4 Component 1 CUP B53D23004750006 and C53D2300080006.
FC acknowledges support from a UKRI Frontier Research Guarantee Grant (PI Cullen; grant reference EP/X021025/1). CK acknowledges funding from the UK Science and Technology Facility Council through grant ST/Y001443/1.\\
This work is based in part on observations made with the NASA/ESA/CSA James Webb Space Telescope. The data were obtained from the Mikulski Archive for Space Telescopes at the Space Telescope Science Institute, which is operated by the Association of Universities for Research in Astronomy, Inc., under NASA contract NAS 5-03127 for JWST. These observations are associated with program 1879.\\
\indent Some of the data presented in this paper were obtained from the Mikulski Archive for Space Telescopes (MAST). The specific observations analyzed can be accessed via \url{https://doi.org/10.17909/v7wt-1r79}. STScI is operated by the Association of Universities for Research in Astronomy, Inc., under NASA contract NAS5–26555.\\
Some of the data products presented herein were retrieved from the Dawn JWST Archive (DJA). DJA is an initiative of the Cosmic Dawn Center (DAWN), which is funded by the Danish National Research Foundation under grant DNRF140.

We thank Vasily Belokurov and Stephanie Monty for providing the catalogue of globular cluster stars.

\end{acknowledgements}

\bibliographystyle{aa}
\bibliography{aa58415-25}

\begin{appendix}

\section{Optimization of strong-line diagnostic combinations for determining metallicity}
\label{sec:tests_inversion}

In order to compute gas-phase metallicities for high-redshift galaxies in the absence of any auroral line detections, we tested a suite of empirical strong-line calibrations available in the literature. In particular, we considered the empirical relations presented in \citet{curti_new_2017, Cataldi2025, 2025ApJ...985...24C, 2025arXiv250810099S}. For each calibration set we adopted the published polynomial coefficients to reconstruct the functional forms linking metallicity to the relevant emission-line diagnostics, namely $R3$ and $R2$ as already defined, $O32 = \log(\OIIIopt~\lambda5007 / \OII~\lambda\lambda3727,3729)$, $\tilde{R} = 0.46 R2 + 0.88 R3$, and $Ne3O2 = \log(\NeIII~\lambda3869 / \OII~\lambda\lambda3727,3729)$. 
We then designed an inversion procedure that, for each galaxy, compares the observed diagnostic values with the expected polynomial curves, and determines the metallicity as the value that minimizes the total $\chi^2$ across all the selected diagnostics.

To evaluate the optimal choice of diagnostics, we constructed a pool of possible line ratios and explored all combinations including at least two diagnostics. For each combination, we computed the distribution of $\chi^2$ values across the galaxy sample and quantified the typical goodness of fit by taking the median $\chi^2$. In addition, we calculated the curvature of the $\chi^2$ profiles as an indicator of the robustness of the metallicity determination, and we also evaluated the variance of the inferred metallicity distribution, quantified as the interquartile range $\left[P_{75} - P_{25}\right]$, to avoid solutions where all diagnostics would predict overly clustered values (thus limiting degeneracies in the inversion). During the procedure, we further excluded from the statistics those galaxies whose inferred metallicities yielded a reduced $\chi^2$ above a fixed threshold (e.g. $\chi^2_\nu > 10$). While this inevitably decreases the final number of galaxies with indirect metallicity estimates, it ensures that the retained measurements are more robust and reliable.

We applied the same diagnostic-selection test to the local reference sample. For the low-redshift galaxies, the combination that formally yields the lowest median $\chi^2$ with respect to the \citet{curti_new_2017} calibrations is $R3$ together with $\tilde{R}$. However, the difference relative to the $R2,R3$ combination is very small, amounting to only $\sim0.5$ in the median $\chi^2$ of the sample. At the same time, we find that the $R2,R3$ pair maximizes both the metallicity variance and the curvature of the $\chi^2$ surface, providing slightly stronger leverage in constraining the solution. Given the marginal difference in $\chi^2$ and in order to maintain full methodological consistency with the high-redshift analysis, we therefore adopt the $R2,R3$ combination also for the local sample.

We also performed cross-calibration tests to assess the consistency of metallicity estimates across different empirical relations. We find that the results can vary significantly depending on the adopted calibration. In particular, the \citet{curti_new_2017} relations, which are based on local-Universe samples, tend to yield systematically different metallicities. Similarly, the calibration by \citet{2025ApJ...985...24C} shows limitations in certain regions of parameter space due to extrapolation issues. These comparisons highlight that the choice of calibration curve is a major source of systematic uncertainty in strong-line metallicity determinations at high redshift.

\section{The origin of the scatter in the N/O calibrations}
\label{sec:scatter_NO_strong_lines}

\begin{figure}
    \centering
    \includegraphics[width=1\linewidth]{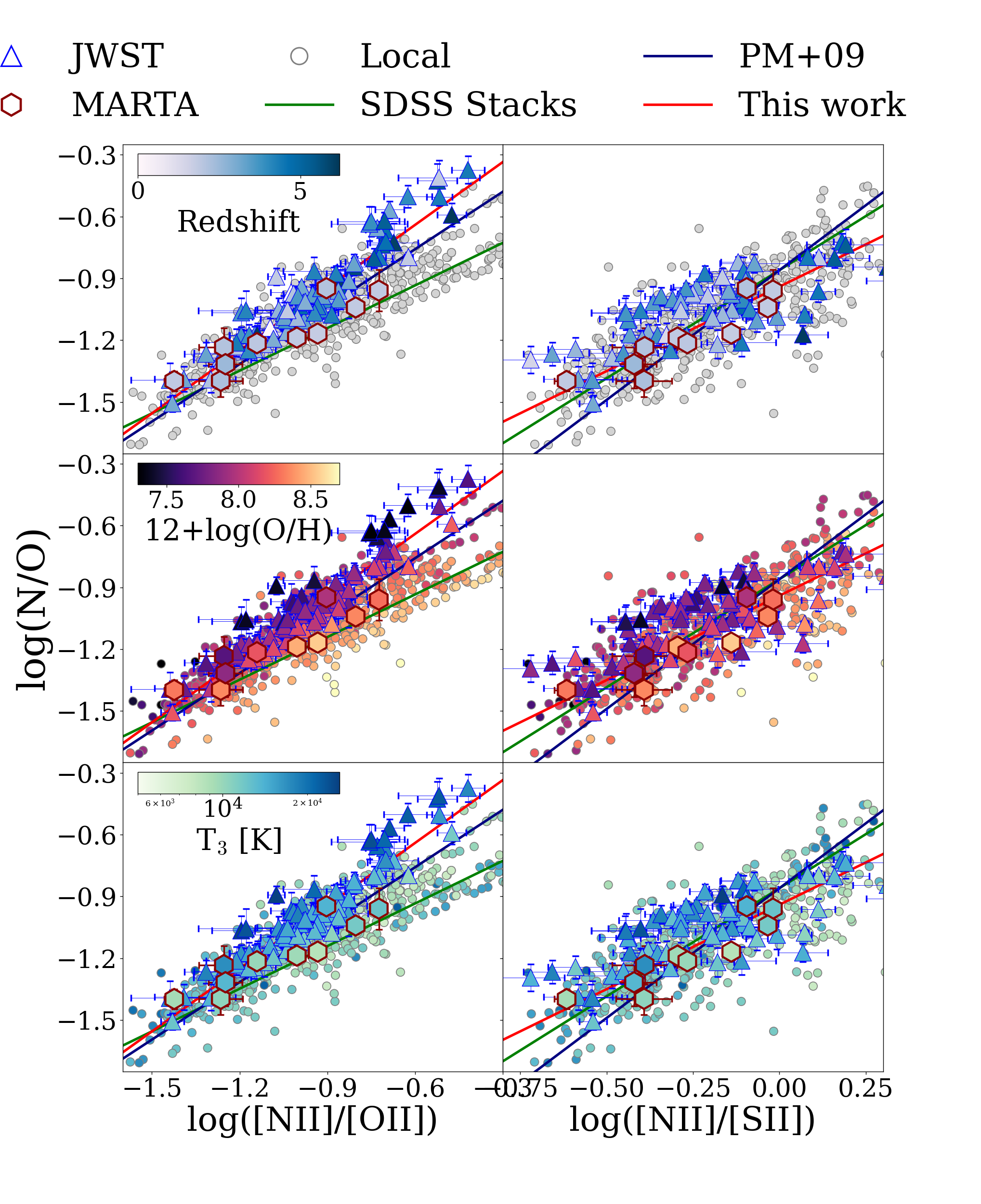}
    \caption{Same as Figure~4, but here we include six panels arranged in three rows by two columns, each showing the relation between directly measured N/O and the observed emission-line ratios \NII/\OII\ (left column) and \NII/\SII\ (right column). Each row adopts a different colour coding: redshift (top), gas-phase metallicity derived via the direct method (middle), and \OIIIopt\ electron temperature, $T_3$ (bottom).}
    \label{fig:NO_trend_all}
\end{figure}

We investigate the physical origin of the intrinsic scatter affecting the empirical N2O2–N/O and N2S2–N/O relations. In particular, we analyse the residuals from the best-fit trends as a function of key physical parameters- redshift, metallicity, and electron temperature \(T_3\)- as well as the local gradients of these quantities across the diagnostic planes. The visual trends are summarized in Figure~\ref{fig:NO_trend_all}, where the three panels on the left refer to the N2O2 diagnostic and those on the right to N2S2; from top to bottom, the colour coding traces redshift, metallicity, and electron temperature.

No clear or systematic trend is found with redshift. The two gradient angles, $\sim 67^\circ$ for N2O2 and $\sim -14^\circ$ for N2S2, have large uncertainties and are therefore not particularly informative. We performed a Monte Carlo permutation analysis in which galaxy redshifts were randomly shuffled while keeping the N/O ratios and line ratios fixed. The resulting $p$-value ($p_{\mathrm{perm}} \approx 0.07$) confirms that any dependence on redshift is statistically marginal. Consistently, the correlations between residuals and redshift are weak and not significant (N2O2: Pearson $r = 0.11$, $p = 0.28$; N2S2: $r = -0.18$, $p = 0.10$). These results indicate that the scatter in both diagnostics does not depend significantly on redshift within our sample ($z \simeq 1.5\text{--}6$), suggesting that the N2O2--N/O and N2S2--N/O relations are not subject to strong evolutionary effects with cosmic time.

Among the parameters considered, the gas-phase metallicity shows the clearest dependence. In both diagnostics, metallicity gradients are oriented along directions intermediate between the diagnostic and the \(\log(\mathrm{N/O})\) axis (\(\theta_Z \approx -48^\circ\) for N2O2 and \(-37^\circ\) for N2S2), and the residuals from the best-fit relations show a strong anti-correlation with metallicity (Pearson \(r = -0.75\) and \(-0.39\) for N2O2 and N2S2, respectively). Specifically, the fit tends to underestimate \(\log(\mathrm{N/O})\) at low metallicity. 

The gradients with electron temperature are similar but show weaker correlations with the residuals (Pearson values of +0.20 and +0.23 respectively), with higher \(T_3\) values corresponding to lower N/O at fixed diagnostic ratio, going in the opposite direction as the trend with metallicity. 

The diagnostics dependence on metallicity can be interpreted as a secondary effect of the ionisation parameter \(U\), which is known to anti-correlate with metallicity \citep[e.g.][]{dopita_evans_1986, perez-montero_deriving_2014, Thomas2019}. Since the \OII\ and \SII\ lines are sensitive to the ionisation structure, variations in \(U\) shift the position of galaxies in both diagnostic planes: at fixed N/O, a decrease in \(U\) moves galaxies toward higher N2O2 or N2S2 values.
Furthermore, recent photoionisation models by \citet{Martinez2025} predict a metallicity dependence of these diagnostics, consistent with the trends observed in our data.

Overall, the analysis shows that the main driver of the scatter in both N2O2 and N2S2 is the secondary dependence on ISM conditions, particularly the interplay between metallicity and ionisation parameter.

\begin{table}[t]
\centering
\caption{Summary of statistical tests for the N/O calibrations.}
\begin{tabular}{lcccc}
\hline
\hline
Relation & Quantity & Angle [deg] & Pearson $r$ & $p$-value \\
\hline
\multirow{3}{*}{\shortstack{N2O2 \\–N/O}} 
 & $z$ & $67 \pm 49$ & $+0.11$ & $0.28$ \\
 & O/H & $-48 \pm 1$ & $-0.75$ & $<10^{-4}$ \\
 & $T_3$ & $136 \pm 14$ & $+0.20$ & $<10^{-4}$ \\
\hline
\multirow{3}{*}{\shortstack{N2S2 \\–N/O}}
 & $z$ & $-14 \pm 26$ & $-0.18$ & $0.10$ \\
 & O/H & $-37 \pm 2$ & $-0.39$ & $<10^{-4}$ \\
 & $T_3$ & $148 \pm 4$ & $+0.23$ & $<10^{-4}$ \\
\hline
\end{tabular}
\tablefoot{Pearson correlation coefficients refer to the residuals of the best-fit relations.}
\label{tab:gradients}
\end{table}

\section{Statistical tests on the N/O-O/H residual distributions}
\label{sec:statistics}

To rigorously quantify the differences between the high and low-redshift distributions in the N/O-O/H plane, and in particular their behaviour relative to the local best-fit relation, we performed a series of statistical tests whose results are summarised in Table~\ref{tab:statistics}. The median offset between the two residual distributions is $0.18$~dex, with a 95\% bootstrap confidence interval of [0.14, 0.20]~dex, which excludes zero at the $\sim10\sigma$ level and therefore confirms the systematic nature of the enhancement.

To further assess the significance of this difference, we applied two complementary non-parametric tests. The AD yields $p < 10^{-3}$, indicating that the two distributions are inconsistent with being drawn from the same parent population. The Mann-Whitney $U$ test\footnote{Also known as the Wilcoxon rank-sum test, the Mann-Whitney test evaluates whether two independent samples differ in their median values. It is based on ranking all data points and comparing the sum of ranks between groups, and does not assume normality.} returns $p \ll 10^{-4}$, further confirming that high-redshift galaxies exhibit systematically larger residuals than their local counterparts.

\begin{table*}
\centering
\caption{Statistical comparison between local and high-$z$ N/O populations.}
\label{tab:statistics}
\begin{tabular}{lccc}
\hline\hline
Property & Local Universe & High-$z$ & Notes \\
\hline
\multicolumn{4}{c}{Sample sizes} \\[2pt]
Total sample & 88,201 & 664 & Combined: 88,865 \\
$12+\log(\mathrm{O/H})<8.1$ & 234 & 121 &  \\ 
$12+\log(\mathrm{O/H})\ge 8.1$ & 87967 & 543 &  \\[4pt]
\hline

\multicolumn{4}{c}{Raw $\log(\mathrm{N/O})$ distribution (no model correction)} \\[2pt]
Median (all) & $-0.923$ & $-1.067$ & $\Delta = +0.144$ dex \\
Std. deviation & $0.195$ & $0.226$ & High-$z$ more scattered \\[4pt]
\hline

\multicolumn{4}{c}{Metallicity-binned comparison} \\[3pt]

$12+\log(\mathrm{O/H})<8.1$ & & & \\
Median residual & $+0.03$ & $+0.35$ & $\Delta = 0.32$ dex \\
Bootstrap 95\% CI & \multicolumn{2}{c}{[0.27, 0.39]} & Excludes zero \\
MW / AD tests & \multicolumn{2}{c}{$p < 10^{-3}$} & Highly significant \\[6pt]

$12+\log(\mathrm{O/H})\ge 8.1$ & & & \\
Median residual & $+0.001$ & $+0.134$ & $\Delta = 0.13$ dex \\
Bootstrap 95\% CI & \multicolumn{2}{c}{[0.111, 0.147]} & Excludes zero \\
MW / AD tests & \multicolumn{2}{c}{$p < 10^{-3}$} & Highly significant \\
\hline
\end{tabular}
\tablefoot{Tests are computed both on the whole sample and in two metallicity bins 12+log(O/H) $<8.1$, and 12+log(O/H) $\geq 8.1$}
\end{table*}

\section{Observational biases potentially affecting the N/O abundances}
\label{sec:N_enhanced}

\begin{figure*}
    \centering
    \includegraphics[width=\linewidth]{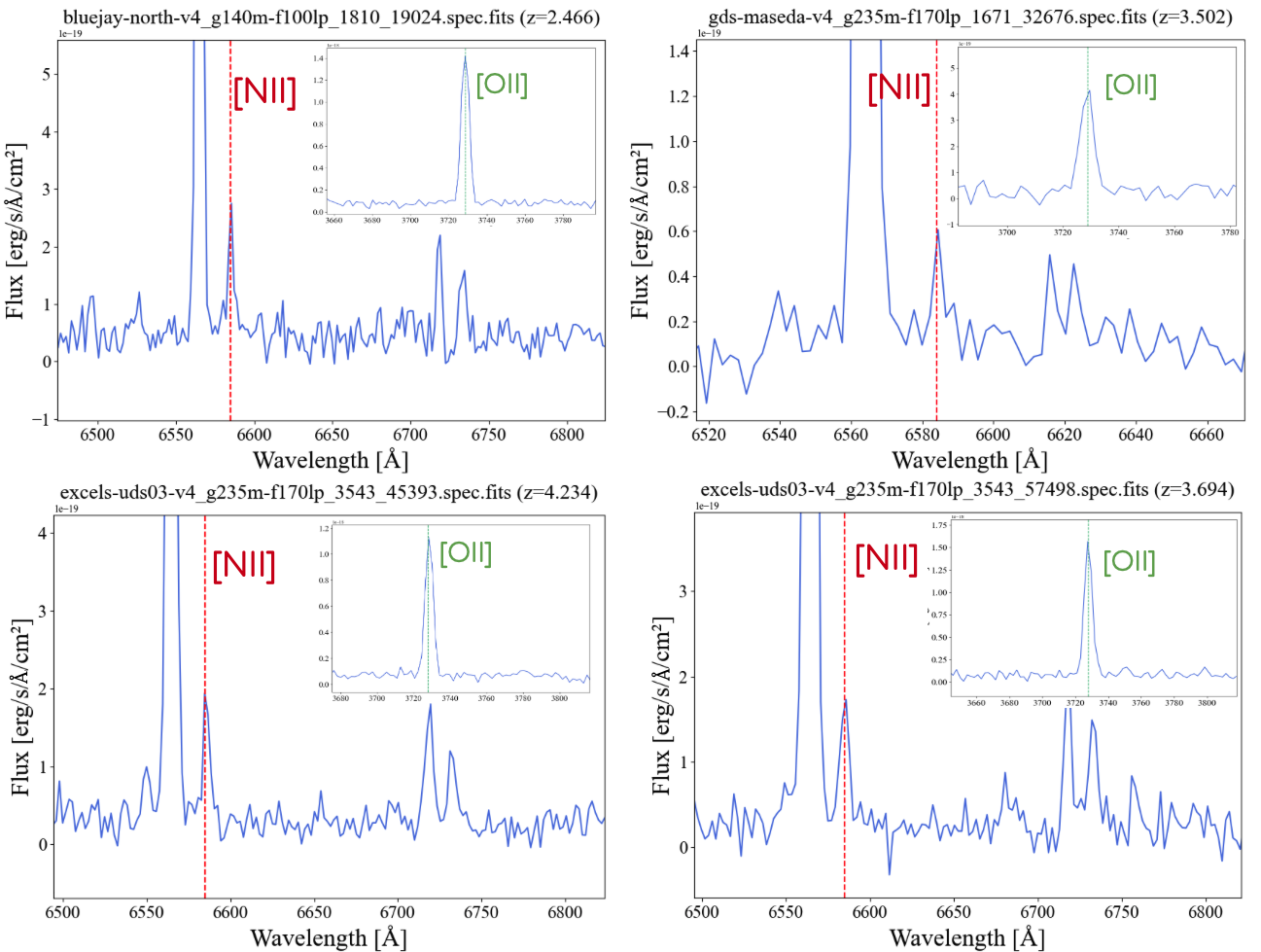}
    
    \caption{Example spectra of N-enhanced galaxies from the DAWN Archive. The \NII\ feature is marked by a red dashed vertical line, while the inset panels highlight the \OII\ detections with green dashed lines. In all cases, both lines are robustly detected with signal-to-noise ratios greater than 3.}

    \label{fig:spectra_N_enhanced}
\end{figure*}

\begin{figure}
    \centering
    \includegraphics[width=1\linewidth]{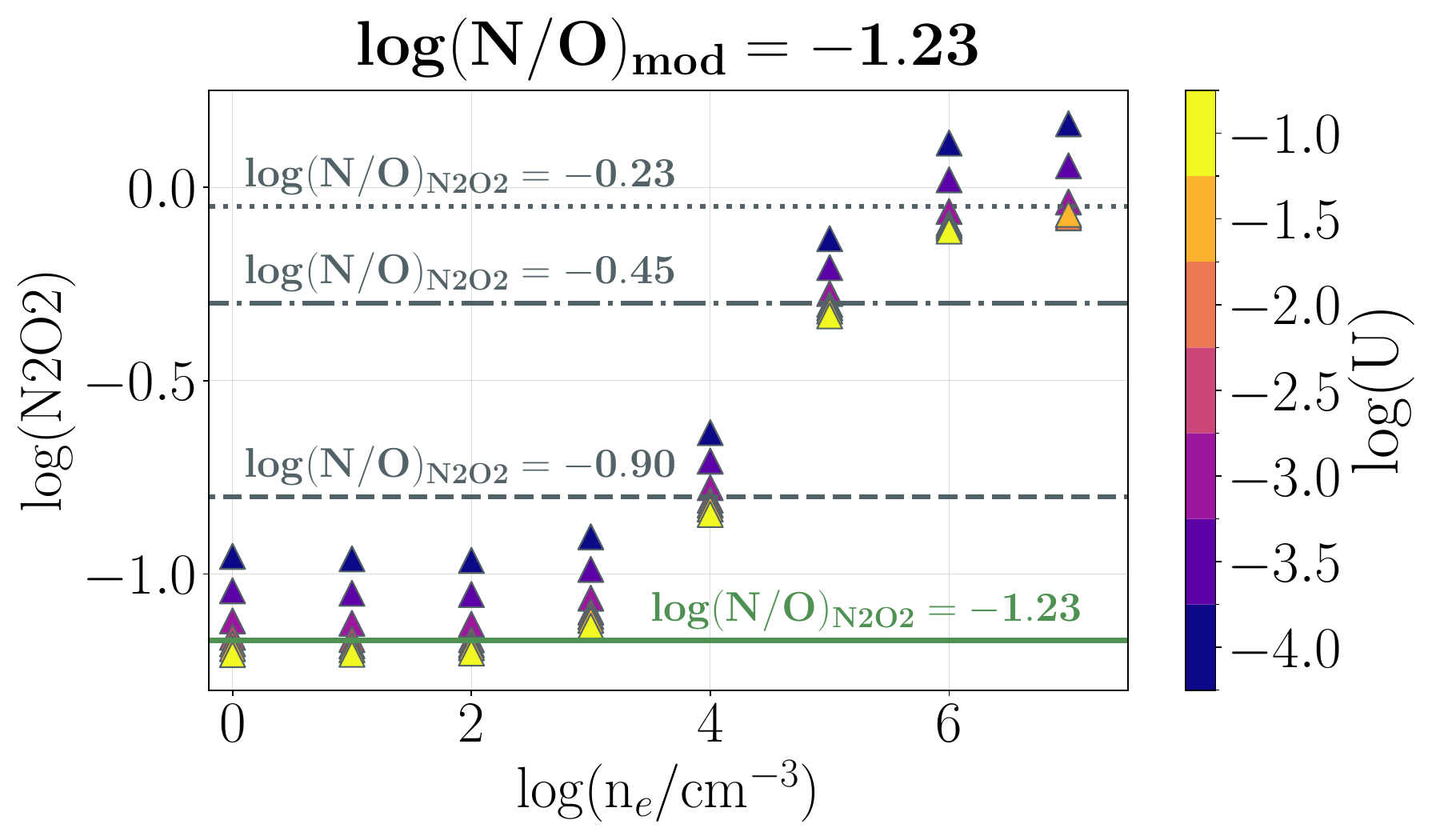}
    \caption{N2O2 as a function of density for the CLOUDY models described in Appendix \ref{sec:N_enhanced} (triangles), colour-coded by the ionisation parameter log(U). The green solid line shows the N2O2 value corresponding to the model N/O using the calibration \ref{eq:n2o2_calibration}. The grey dashed, dot-dashed, and dotted lines indicate the N/O that would be inferred from the calibration at those specific N2O2. While our calibration performs well at densities $\leq10^3$ cm$^{-3}$, it overestimates N/O at higher gas densities, as expected.}
    \label{fig:N2O2_density}
\end{figure}

We consider how selection effects may be driving the position of high-$z$ galaxies in the N/O-O/H plane. 
To further investigate the population of low-metallicity, nitrogen-enhanced galaxies identified in Sect. \ref{sec:no_vs_oh}, we performed a detailed inspection of their spectra. As a first step, we carried out a visual examination of the \NII\ and \OII\ line detections to ensure their reliability. All galaxies classified as N-enhanced exhibit reliable detections of both lines, although a small subsample (three to four objects) shows marginal detections with S/N~$\sim$3. Figure~\ref{fig:spectra_N_enhanced} displays four representative examples of N-enhanced galaxies from the DAWN Archive, all selected to have metallicities below 12+log(O/H)$<8$ and log(N/O)$>-1$. In all cases, the detections are robust.

It is also important to consider whether detection biases could affect the identification of N-enhanced galaxies, particularly in light of the recent findings by \citet{Zhu2025}. These authors showed that, in $z>5$ JWST spectra, the detectability of UV nitrogen lines (\NIII$\lambda\lambda$1747,1749 and \NIV$\lambda1486$) is strongly limited by current observational depths. However, these constraints do not directly apply to our analysis. In contrast to the faint UV lines considered by \citet{Zhu2025}, our classification relies on bright rest-frame optical emission lines such as \NII, \OII; 
furthermore, a large fraction of our sample comes from some of the deepest JWST spectroscopic programs, and at redshift mostly between 2 and 4.

To test whether detectability nonetheless influences the observed N/O–O/H distribution, we examined the S/N of the \NII\ line within our sample. However, we do not observe any clear correlation between S/N and the measured N/O values. Galaxies with strongly enhanced N/O include both weaker detections (S/N~$\sim$3–5) and very secure ones (S/N~$>$10). This seems to indicate that detectability alone is unlikely to drive the entire observed population of N-enhanced systems. In particular, several JADES galaxies exhibit \NII\ detections with S/N~$>$15; even if their intrinsic \NII\ fluxes were a factor of $\sim$5 lower (i.e., consistent with a “normal’’ N/O ratio), they would still remain detectable within the same observations. The scarcity of such objects therefore suggests that the excess of high N/O systems is not solely an observational artifact but reflects an underlying physical trend.
To further assess whether the observed N/O–O/H trend could be driven by our S/N selection on the \NII\ line, we extended the analysis to include galaxies without a formal \NII\ detection (i.e. S/N$<3$). The concern is that requiring a detection at low metallicity might preferentially select systems with intrinsically enhanced nitrogen emission, thereby artificially biasing the relation toward high N/O values.

If this were the dominant effect, we would expect that most galaxies lacking a reliable \NII\ detection would lie along the local N/O–O/H relation, with only the most extreme N-enhanced systems (i.e. those significantly above the local relation) remaining detectable. To test this scenario, we recomputed N/O and O/H for the non-detected sources using the same methodology adopted throughout the paper: direct-method abundances for the subset with robust auroral-line measurements (X objects), and strong-line calibrations for the remaining sources (Y objects). For galaxies with S/N$<3$ in \NII, we treated the measured fluxes as $3\sigma$ upper limits and propagated them consistently into the abundance determination.

\begin{figure}
    \centering
    \includegraphics[width=\linewidth]{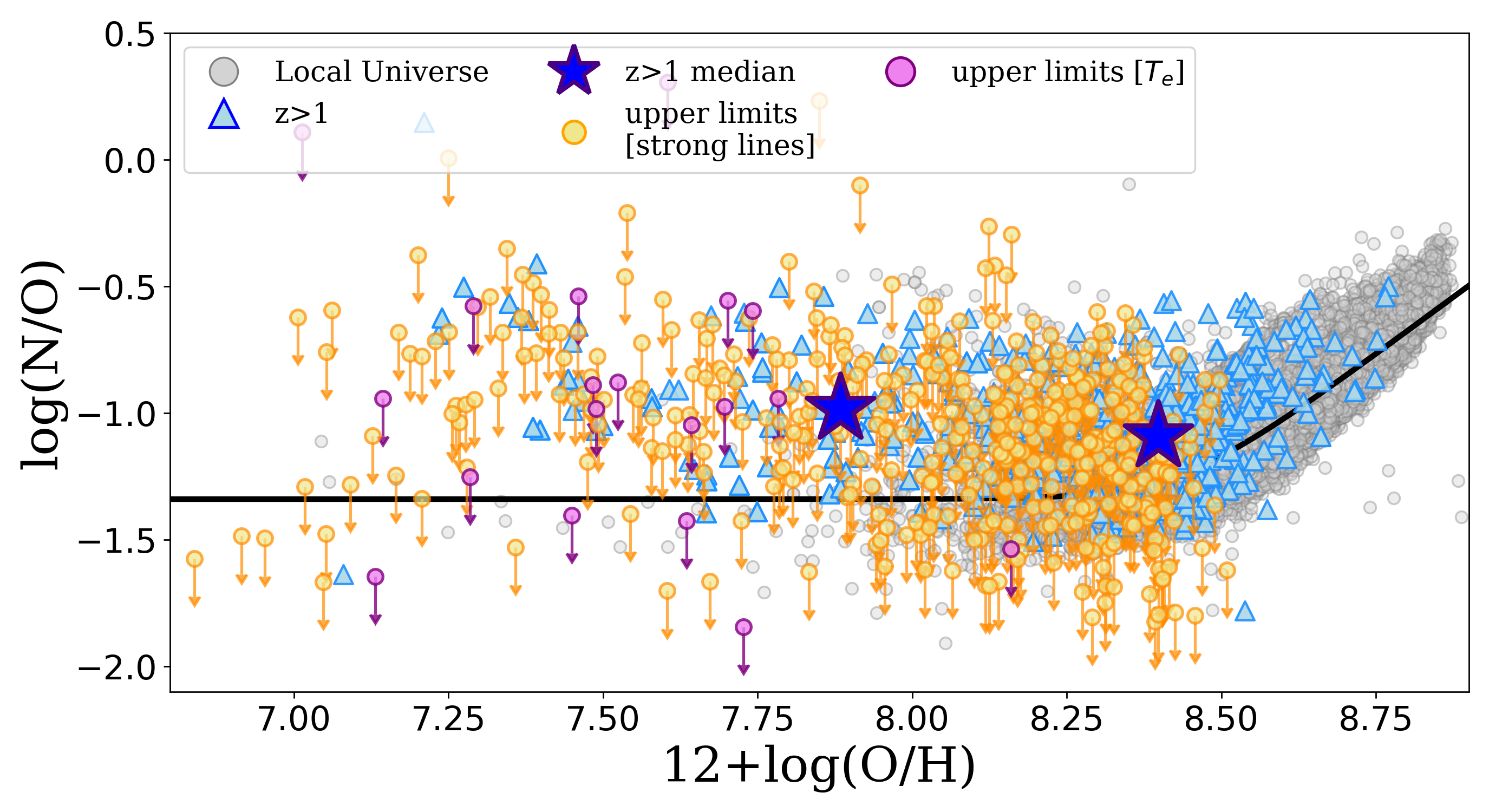}
    
    \caption{Same as Figure \ref{fig:NO_metallicity_bestfit}, but including upper limits on \NII, shown for direct measurements (purple) and strong-line estimates (orange).
}

    \label{fig:upper_limits}
\end{figure}

The results are shown in Figure~\ref{fig:upper_limits}. We do not observe a clear tendency for the upper limits to align with the local N/O–O/H relation. Instead, both direct- and strong-line-based upper limits populate a region broadly consistent with that defined by the detected sources, and their distribution appears compatible with the main observed trend. While this does not directly contradict our conclusions, we acknowledge that the presence of upper limits could still be hiding a lower underlying trend.
For this reason, we are currently developing a forward model aimed at reproducing the ``median'' high-redshift galaxy population, which will be presented in forthcoming work. 

As an additional check, we quantified how many of the N-enhanced galaxies would still satisfy our S/N$>3$ requirement on \NII\ if they were instead placed on the local N/O–O/H relation. For each object, we computed a ``reduction factor'', defined as the factor by which its observed \NII\ flux would need to be decreased in order to match the local fit. We then evaluated whether the rescaled flux would remain above our detection threshold. We find that $\sim$43\% of the galaxies would still be detected after applying this reduction. This result indicates that almost half of the enhanced sample would remain observable even if their intrinsic N/O ratios were consistent with the local relation. While this does not constitute a confirmation of our results, it at least rules out the possibility that the observed N-enhanced population is entirely driven by objects lying just at the detection limit. Furthermore, based on this result, if the local relation were the typical condition for high-z objects, we would expect to observe a non-negligible fraction of galaxies distributed along the local relation, since they would still be observable, but this is not the case.

Beyond detectability considerations, other observational effects may also influence the observed trend. In particular, enhanced N/O ratios have been shown to emerge from density effects arising from compact, unresolved high-density substructures within galaxies \citep{2023ApJ...957...77P, 2024MNRAS.535..881J, Martinez2025, Arellano-Cordova25b, Ji2026}. The critical density of \NII$\lambda$6584 is $\sim10^5
\rm cm^{-3}$ while \OII has a significantly lower critical density of $\sim10^3 \rm cm^{-3}$. In the presence of dense clumps, the \OII flux is therefore suppressed relative to \NII, thereby artificially elevating the inferred N/O ratio.
For example in Figure \ref{fig:N2O2_density} we show the dependence of the N2O2 ratio as a function of density as computed with CLOUDY from single-cloud models; this is indeed an extreme simplification but this is just to show what would happen in presence of very high densities to this diagnostic. For the models, we consider the star-forming models described in \citet{Ceci25}, which span a wide range of densities (log(n$_e$) = [0, 7]) and ionisation parameters (log U = [-4, -1]). These models are computed with CLOUDY version 23.01 \citep{Gunasekera23}, adopting as ionising continua BPASS stellar population models with binaries \citep{Stanway_Eldrige_BPASS_2018} and a \citet{Kroupa2002} IMF with an upper mass cutoff of 300 M$\odot$.  
For this analysis we adopt a fixed stellar age of \(10^6\) yr, a stellar metallicity of \(\log(Z_\star) = -1.7\), and no depletion onto dust grains. All models assume a solar abundance pattern from \citet{Asplund21} for all elements, except for carbon and nitrogen, which are rescaled following the \citet{nicholls_abundance_2017} relations, with an additional +0.2 dex offset applied to nitrogen.

As visible in Figure~\ref{fig:N2O2_density}, the  ratio becomes increasingly biased toward higher values as the gas density rises, and a mild vertical dependence on the ionisation parameter is also present, although it remains subdominant relative to the density effect. While it is true that real galaxies host a far more complex, multiphase ISM, the models demonstrate that very dense, unresolved substructures could have a non-negligible impact on the observed  ratio.

However, robust diagnostics of such extreme densities, such as UV doublet ratios of highly ionised species (e.g., \CIII, \NIII, \SiIII, \NIV), are not available for large high-z samples. Even when these lines are detected, it is important to note that such ions do not necessarily trace gas that is co-spatial with the regions emitting \NII and \OII \citep[e.g.][]{kewley_understanding_2019, Choustikov2026}. In a simplified two-component picture, if the high-density component is also the brightest--i.e. it dominates the emergent line flux--then the density inferred from the UV transitions may reasonably approximate the physical conditions relevant for \NII as well. Conversely, if the low-density component contributes substantially to the total flux, a more sophisticated multi-phase treatment would be required to correctly assess the bias affecting N2O2. Therefore, with current data, we cannot yet precisely quantify the extent to which density effects contribute to the observed N/O enhancement.

Finally, we explored how these N-enhanced galaxies behave with respect to the star-forming main sequence, in order to assess if we are only considering starburst objects in our sample or if we are probing a fair fraction of the ``normal'' galaxy population. 

\begin{figure}
    \centering
    \includegraphics[width=\linewidth]{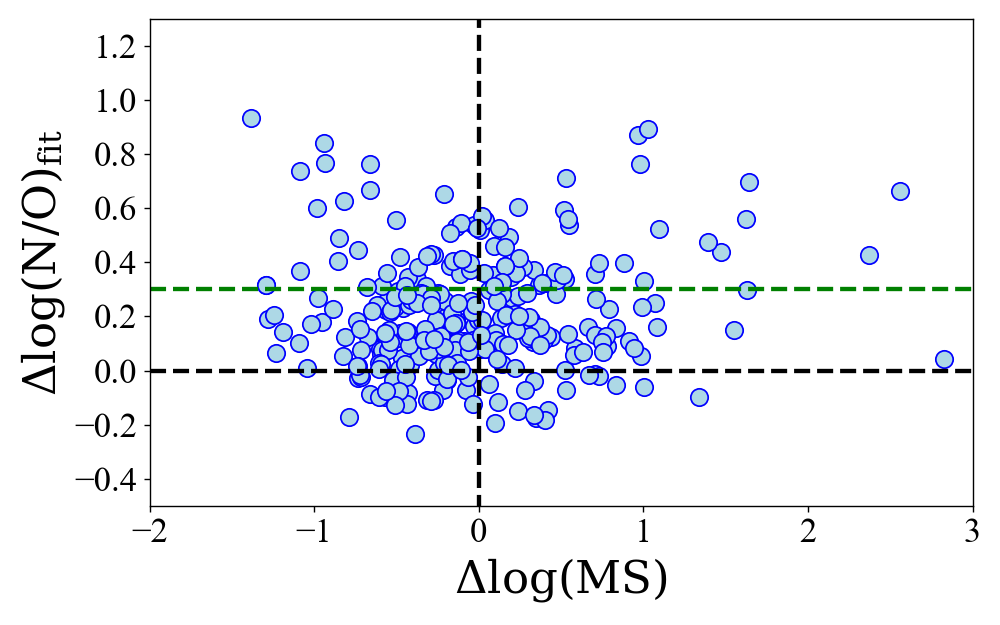}
    
    \caption{Representation of the offset $\Delta \log(\mathrm{N/O})$ of the high-redshift data with respect to the local $\mathrm{N/O}$ best-fit relation, as a function of $\Delta_{\mathrm{MS}}$, i.e. the offset of the same galaxies from the MS of \citet{popesso_SFMS_2023}, computed at the specific redshift of each object. The black dashed lines indicate the zero offsets, corresponding to perfect agreement between the observed $\log(\mathrm{N/O})$ and the local best-fit relation, and between the observed SFR and the SFR predicted by the Popesso et al. Main Sequence. The green dashed line marks the adopted enhancement threshold, corresponding to galaxies with $\log(\mathrm{N/O})$ higher by more than 0.3 dex than expected. As visible from the figure, no clear trend of the $\mathrm{N/O}$ enhancement with $\Delta_{\mathrm{MS}}$ is observed.
}

    \label{fig:delta_MS}
\end{figure}

The results are shown in Figure~\ref{fig:delta_MS}, where we plot the offset $\Delta \log(\mathrm{N/O})$ from the local best-fit relation as a function of $\Delta_{\mathrm{MS}}$, i.e. the deviation from the star-forming main sequence of \citet{popesso_SFMS_2023} computed at the redshift of each object. The black lines mark zero offsets in both quantities, while the green dashed line indicates the adopted threshold for nitrogen enhancement, $\Delta \log(\mathrm{N/O}) > 0.3$.

An inspection of Figure~\ref{fig:delta_MS} shows no clear trend between $\Delta \log(\mathrm{N/O})$ and $\Delta_{\mathrm{MS}}$. Nitrogen-enhanced galaxies are distributed across the full range of offsets from the main sequence, including systems lying above, on, and below the MS. $\sim 29\%$ of the galaxies in our sample are classified as N/O-enhanced. Among these, the $\sim 32\%$ lie above the MS ($\Delta_{\mathrm{MS}} > 0.3$), $\sim 39\%$ are consistent with the MS ($-0.3 < \Delta_{\mathrm{MS}} < 0.3$), and $\sim 29\%$ fall below it ($\Delta_{\mathrm{MS}} < -0.3$). This relatively even distribution indicates that nitrogen enhancement is not confined to starburst systems, but is also common among galaxies with more typical star-formation rates.

A more comprehensive investigation of the interplay between stellar mass, metallicity, SFR, and nitrogen enrichment requires a multi-parameter approach that accounts for additional physical properties. 
A detailed analysis incorporating multiple properties, along with a careful reassessment of stellar mass estimates across the full sample, will be presented in a forthcoming study.

\end{appendix}

\end{document}